\documentclass[preprint,5p,times,twocolumn]{elsarticle}
\usepackage{framed} 
\usepackage{multicol} 
\usepackage{nomencl} 
\makenomenclature
\usepackage{tabularx}
\renewcommand*\nompreamble{\begin{multicols}{2}}
\renewcommand*\nompostamble{\end{multicols}}
\usepackage{amsmath}
\usepackage{amssymb}
\usepackage{arydshln} 
\usepackage{amsthm}
\usepackage{algorithm}
\usepackage{algorithmic}
\usepackage{graphicx}
\usepackage{subfig}
\usepackage{color}
\usepackage{url}
\definecolor{lightblue}{RGB}{0,176,240}
\definecolor{purple}{RGB}{153,102,255}
\usepackage{amsmath}
\usepackage[scr=boondox,  
            cal=esstix]   
           {mathalpha}
\usepackage{algorithm}
\usepackage{float}
\usepackage{multirow}
\usepackage[utf8]{inputenc}
\usepackage[colorlinks,linkcolor=blue, anchorcolor=blue, citecolor=blue]{hyperref}
\usepackage[superscript,nomove,biblabel]{cite}
\usepackage[switch]{lineno}
\makeatletter
\def\ps@pprintTitle{%
 \let\@oddhead\@empty
 \let\@evenhead\@empty
 \let\@oddfoot\@empty
 \let\@evenfoot\@empty
}
\makeatother
\journal{}
\begin{document}

\begin{frontmatter}
\title{Long-range near-surface wake signatures of offshore wind farm clusters revealed by satellite observations}

\author[label1]{Rui Li}
\ead{lironui@outlook.com}
\author[label1]{Jincheng Zhang}
\address[label1]{Intelligent Control \& Smart Energy (ICSE) Research Group, School of Engineering, University of Warwick, Coventry, CV4 7AL, UK}
\ead{jincheng.zhang.1@warwick.ac.uk}
\author[label1]{Xiaowei Zhao\corref{cor2}}
\ead{xiaowei.zhao@warwick.ac.uk}
\cortext[cor2]{Corresponding author}

\begin{abstract}
Wind energy has expanded rapidly in recent years, leading to increasingly dense clusters of offshore wind farms. As a result, wind farm wake effects, manifested as reduced wind speeds downstream of operating turbines, have become an important consideration for wind resource assessment and large-scale planning. Here we investigate near-surface wake signatures using wind speeds retrieved from 7,122 Sentinel-1A/B synthetic aperture radar (SAR) images acquired between 2020 and 2022, covering more than 60 offshore wind farms across Europe and Asia. A consistent processing workflow and an automated wake-detection algorithm are applied to identify wake-affected regions and quantify wake-related wind speed changes at 10 m height. The results show that near-surface wake signatures can persist over distances exceeding 100 km under favorable conditions and that wake-affected regions exhibit an average wind speed reduction of 0.990 m/s (12.4\%) at 10 m height. Several cases of wake interactions extending across national boundaries are also observed in densely developed offshore regions. These findings provide a large-scale, observation-based characterization of near-surface wind farm wake signatures.
\end{abstract}

\end{frontmatter}
\setcounter{secnumdepth}{0}


\section{Introduction}

Wind energy is under sustained and accelerated development in recent years, driven by the soaring energy prices across the globe and the imminent net-zero target to tackle climate change \cite{perera2023challenges, howland2022collective, cherp2021national}. Compared with their onshore counterparts, offshore wind farms offer numerous advantages, including higher wind velocities with reduced turbulence, enhanced social acceptance, greater space availability, and heightened energy density owing to the feasibility of deploying larger turbines\cite{pettas2021effects}. When selecting an appropriate site for constructing a new wind farm, various design factors should be considered, including wind resource potential, initial construction cost, operation and maintenance expenses and proximity to transmission lines and roads \cite{lundquist2019costs, li2025flowformer}. However, finding sea areas that meet all these requirements can be challenging. Consequently, offshore sites with favorable wind conditions, soil quality and water depth, such as the North Sea and the Yellow Sea, are witnessing the proliferation of closely spaced wind farms \cite{xu2020proliferation, costoya2021climate}. For example, dozens of large-scale wind farms owned by different countries are already operating in the North Sea, with more farms planned for the future \cite{pryor2024wind}. As these sea areas become increasingly crowded, it is crucial to consider and quantify the wake effects, which refer to the reduction in wind speeds in the downstream region caused by upstream wind farms. Understanding and addressing these wake effects is essential to navigate legal issues and conflicts that will inevitably arise from uncoordinated wind farm development in the foreseeable future, thereby enhancing the sustainability in wind energy planning and utilization \cite{finseraas2022gone, duvivier2020moat}. 

Wind farm wakes, as a complex aerodynamic phenomenon, are influenced by many factors such as upstream wind speeds, atmospheric stability and meteorological conditions. Currently, two mainstream technologies are used to investigate the potential impacts of wake effects: numerical simulations \cite{fischereit2022review, larsen2021case, siedersleben2018micrometeorological, fitch2012local, ali2023assessment} and in-situ measurements \cite{platis2018first, lampert2020situ, barfuss2020airborne, sebastiani2022evaluation, schneemann2021offshore}. For example, high-resolution microscale and urban climate models, such as Parallelized Large-Eddy Simulation Model (PALM) \cite{maronga2020overview, vollmer2015first, witha2014high, witha2014large, dorenkamper2015impact} and Microscale Transport and Stream Model (MITRAS) \cite{salim2018microscale, schlunzen2018scientific, boettcher2015influence}, have demonstrated the capability to resolve wind farm wake dynamics and turbine-induced flow modifications at fine spatial scales. At the same time, a series of in-situ measurement campaigns \cite{lissaman1983wake, lee2017observing, platis2018first, barfuss2020airborne}, including mast- and LiDAR-based observations, have provided direct observational evidence of wake effects in both onshore and offshore environments. However, numerical simulations cannot encompass the full range of real-world operating conditions, as they are constrained by prescribed boundary conditions and parameterizations. Conversely, in-situ measurements are inherently limited in spatial coverage, restricting their ability to capture wake behavior across multiple wind farms and broad geographic regions. Although previous observational and modeling studies have documented wake extensions exceeding 50–70 km under certain atmospheric conditions \cite{howland2022collective, platis2018first, djath2018impact}, a systematic, multi-site, observation-based assessment of the statistical prevalence and spatial extent of near-surface wake signatures across large offshore wind farm clusters remains limited. This creates a gap between detailed local-scale studies and large-scale, observation-based characterization of near-surface wind-field modifications induced by wind farms, which satellite observations are uniquely positioned to address.

\begin{figure*}[htb]
\centering
\includegraphics[width=16cm]{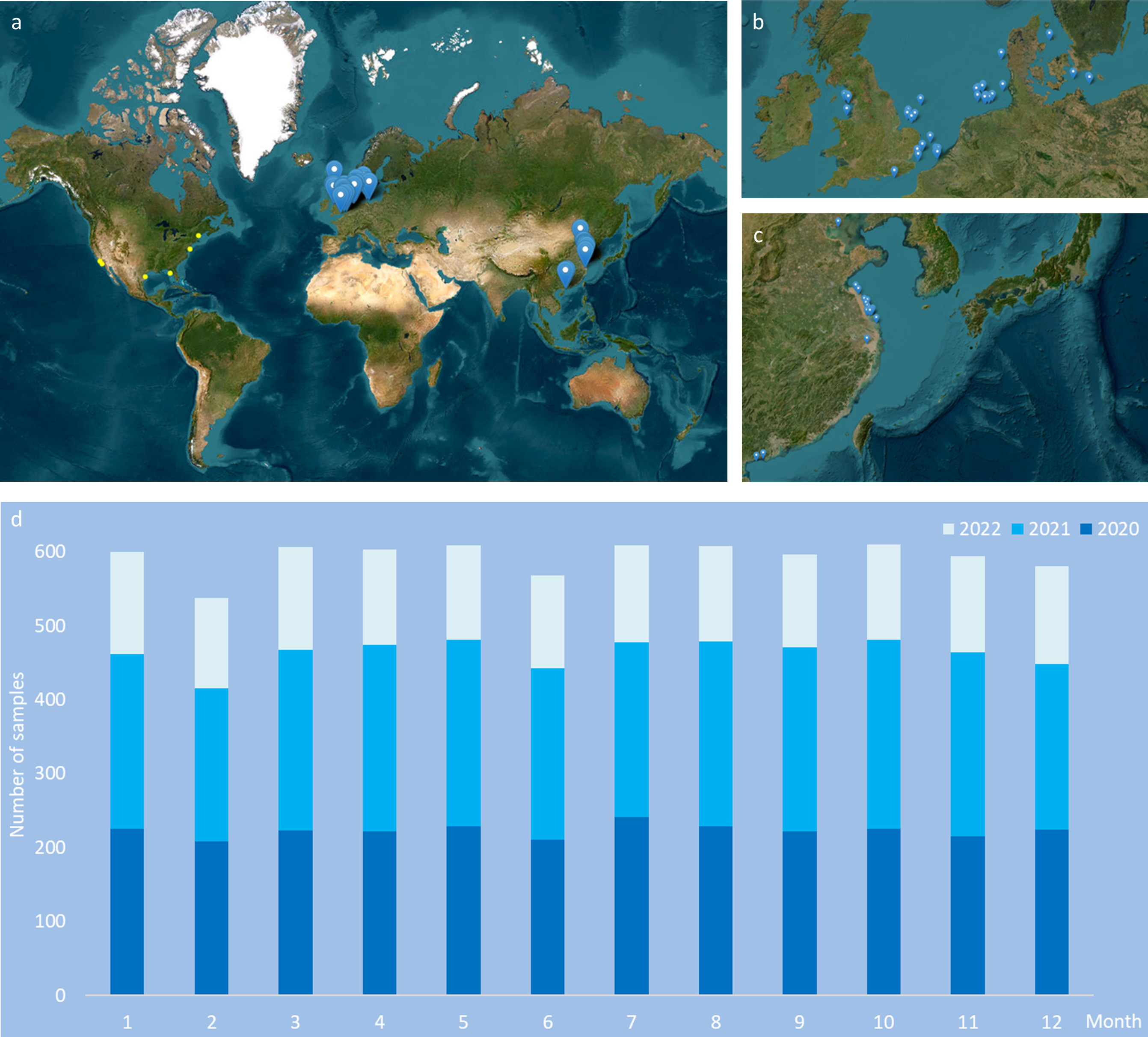}
\caption{\textbf{Spatial and temporal distribution of Sentinel-1A/B SAR acquisitions.} (a) Geographic distribution of wind farms included in this study; yellow markers near the United States indicate buoy locations used for validation of SAR-derived wind speeds. (b) Distribution of European wind farms. (c) Distribution of Asian wind farms. (d) Monthly sample counts from 2020 to 2022. Map Credit: The satellite imagery layer is the intellectual property of ESRI and is used herein under license (Copyright \copyright{} ESRI and its licensors. All rights reserved). The underlying geographic basemap data is sourced from OpenStreetMap and is open access, freely available under the Open Data Commons Open Database License (ODbL) (\copyright{} OpenStreetMap contributors).}
\label{fig:1}
\end{figure*}

This study aims to address the aforementioned questions by utilizing Synthetic Aperture Radar (SAR) images to comprehensively investigate the impact of wind farm wakes. SAR images provide a reliable and universal means to analyze near-surface wind speeds in offshore environments, as the 10 m wind speed can be retrieved from these images using the C-band geophysical model function \cite{hersbach2010comparison}. The reliability and consistency of SAR-derived wind speeds for offshore wind farms under both free-stream and wake conditions have been previously validated \cite{ahsbahs2018applications}. SAR images, captured by satellite-borne sensors, offer the advantage of global coverage, enabling the characterization of near-surface wake signatures worldwide with high fidelity. Although SAR images have been used to investigate wake effects in previous studies \cite{christiansen2005wake, siedersleben2018micrometeorological, nezhad2021new, schneemann2020cluster, ahsbahs2020us, platis2021evaluation, djath2019wind, djath2018impact}, the scale of SAR data and the range of research areas have been limited, leaving the full potential of remote sensing technology untapped for wind energy applications. 

In this work, we collect a comprehensive dataset consisting of 7,122 Sentinel 1A/B SAR images, totaling more than 11.5 TB of data, obtained from the Copernicus Open Access Hub over a three-year period (2020-2022). These images cover over 60 offshore wind farms across Europe and Asia. The spatial and temporal distributions of the wind farms analyzed in this study are illustrated in Fig. \ref{fig:1}. Only SAR images captured after the wind farms were fully commissioned are utilized to ensure a consistent evaluation. Furthermore, the data collection spans evenly all months of the year, as shown in Fig. \ref{fig:1}(d), allowing for the examination of seasonal variations in wake behavior. Based on the collected data, a processing pipeline is developed to retrieve the wind speed while the accuracy is verified by the buoy data. Finally, an image-processing method is proposed to obtain the upstream freestream wind speeds and the downstream wake speeds. 

The wind farm wake effects have profound economic implications globally and are acknowledged as a crucial concern by both academia and industry. However, neither the government nor the wind energy industry has established a settlement or legislation to compensate downstream wind farms for the revenue loss caused by upstream wind farm wakes yet \cite{lundquist2019costs, finseraas2022gone}. This lack of consensus can be attributed, in part, to the absence of a measurable and quantifiable index to assess the real economic impact of wake effects. The evidence provided by this work from SAR-derived wake-induced near-surface wind speed deficit assessment may accelerate the policy-making processes by providing technical support to address such issues. By analyzing thousands of Sentinel-1 SAR scenes across multiple offshore wind farm clusters, this study bridges the gap between isolated case studies and multi-regional statistical confirmation of near-surface wake signatures.

\begin{table*}[htb]
\setlength{\abovecaptionskip}{0.cm}
\caption{The details of the pre-processing procedure for SAR images.}
\label{table:1}
\centering
\begin{tabularx}{\textwidth}{p{4cm} X}
\hline
Procedure                & Details \\ \hline
Thermal Noise Removal    & The backscatter disturbances caused by thermal noise are corrected, generating more seamless SAR images.               \\
Orbit Correction        & The orbital information is updated in the SAR data using a more accurate Orbit State Vector (OSV) file, which is selected as Sentinel Precise in this work.                \\
Radiometric Calibration  & The reflectivity of SAR is converted into physical units of normalized backscatter by normalizing the reflectivity using a reference plane.              \\
Speckle Noise Filtering  & The Refined Lee \cite{lee2008improved} method is adopted to remove speckles and smooth the grainy salt and pepper effect.                   \\
Bright Object detection & The adaptive thresholding function embedded within the SNAP with a Two-Parameter Constant False Alarm Rate (CFAR) Detector is adopted. The target size, target/guard/background window sizes and Probability of False Alarm (PFA) ($ 10^{-x} $) are set as (30, 300), (10, 350, 600) and 12.5.           \\ 
Multi Looking & 50 by 50 multi-looking operation is conducted in this procedure.          \\ \hline 
\end{tabularx}
\end{table*}

\section{Methods}

\subsection{Wind speed retrieved from the SAR images}

\begin{figure*}[htb]
\centering
\includegraphics[width=16cm]{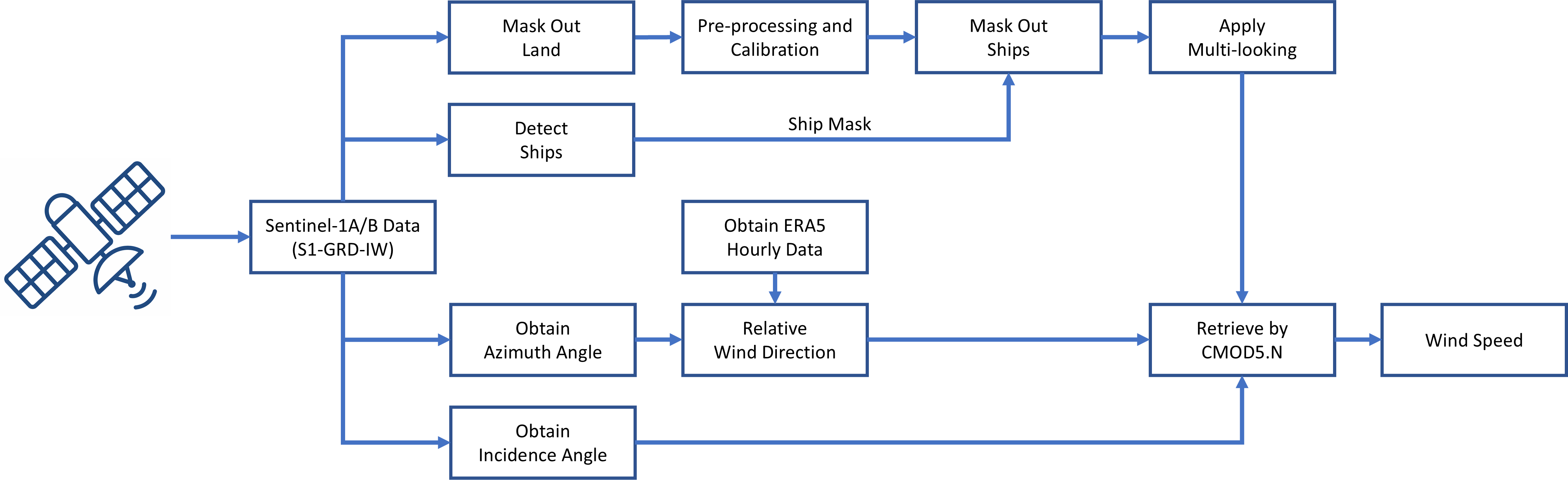}
\caption{\textbf{Workflow of SAR-based wind speed retrieval.} Sentinel-1A/B GRD (IW mode) data are first processed to mask out land and ships, followed by calibration and multi-looking. Ship detection is performed to generate a ship mask for removing contamination. The azimuth and incidence angles are extracted from SAR data. ERA5 hourly data are used to derive the relative wind direction. These parameters are then used as inputs to the CMOD5.N geophysical model function to retrieve near-surface wind speed.}
\label{fig:2}
\end{figure*}

In this study, we utilize the CMOD5.N algorithm \cite{hersbach2010comparison} to retrieve wind speeds from the SAR images. The entire retrieval process is summarized in Fig. \ref{fig:2}. Initially, we collect Sentinel-1A/B SAR images from the Copernicus Data Space Ecosystem (\url{https://browser.dataspace.copernicus.eu/}), with the product type being GRD and the sensor mode set to Interferometric Wide (IW). Subsequently, we apply pre-processing and calibration using the Sentinel Application Platform (SNAP) toolbox (\url{http://step.esa.int/main}), including thermal noise removal, orbit correction, radiometric calibration and speckle noise filtering. Furthermore, we mask out land areas and remove bright objects like turbines and ships, as they tend to create strong radar reflections in the SAR image. The resulting images are then subjected to multi-looking, resulting in a final resolution of 500 meters. For wind speed retrieval using CMOD5.N, three external parameters are required: azimuth angle, incidence angle and wind direction. The first two parameters could be directly obtained from the SAR images themselves. Regarding wind direction, ECMWF Reanalysis version 5 (ERA5) hourly data (\url{https://www.ecmwf.int/en/forecasts/dataset/ecmwf-reanalysis-v5}) \cite{hersbach2018era5} are used as a reference. By combining all these procedures, the wind speed could be eventually retrieved from the SAR images. The details of each pre-processing step are provided in Table \ref{table:1}. 

\subsection{Verification of retrieved wind speed using buoy data}

\begin{figure*}[htb]
\centering
\includegraphics[width=16cm]{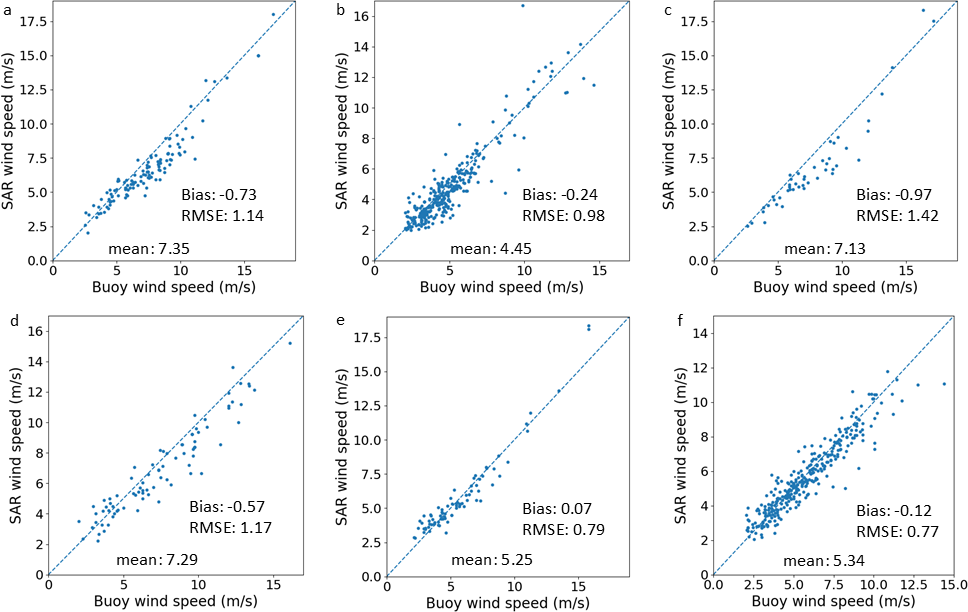}
\caption{\textbf{Comparison of SAR-retrieved wind speed and NDBC buoy wind speed.} Scatter plots show Sentinel-1A/B SAR wind speeds retrieved using CMOD5.N versus buoy wind speeds from the National Data Buoy Center (NDBC). The dashed line denotes the 1:1 relationship. Bias, root mean square error (RMSE), and mean buoy wind speed are shown in each panel. Panels (a–f) correspond to six buoy sites.}
\label{fig:3}
\end{figure*}

To assess the accuracy of the wind speed retrieved from the SAR images, we conduct a verification process using buoy data provided by the National Data Buoy Center (NDBC) (\url{https://www.ndbc.noaa.gov}). We collect data from six different NDBC buoy stations as shown in Fig. \ref{fig:1} (a) and corresponding SAR images from the same locations spanning the period from 2020 to 2022. The buoy station IDs and their respective coordinates are as follows: 42020 (26.968N, 96.693W), 46025 (33.755N, 119.045W), 44009 (38.460N, 74.692W), 44007 (43.525N, 70.140W), 42036 (28.501N, 84.508W) and 46086 (32.499N, 118.052W). Next, we process the SAR images using our developed C-band model pipeline to retrieve the wind speed. We then extract the wind speed values at the corresponding buoy locations based on their coordinates and compare them to the buoy data. In total, we obtain 945 matched pairs of buoy data and retrieved wind speeds (with 117, 304, 46, 84, 77 and 317 pairs for each station, respectively). Since most of the NDBC anemometers are installed at a height of 2.5-4 meters above the sea surface, we convert the measured speeds to the 10-meter wind speed using the logarithmic laws \cite{manwell2010wind}:

\begin{equation}
{u}_{(z_r)} = \frac{{\ln{\frac{z_r}{z_0}}}}{\ln{\frac{z}{z_0}}}{u}_{(z)}, \label{equa:1}
\end{equation}  

\noindent where $ {u}_{(z_r)} $ and $ {u}_{(z)} $ represent the wind speeds at the height $ z_r $ and $ z $, respectively. $ z_0 $ is the surface roughness, which was set to 0.0002 m for the sea surface \cite{charnock1955wind, landberg2003wind}.

The scatter plots depicting the comparison results for the six cases can be seen in Fig. \ref{fig:3}, while detailed statistical metrics are reported in Table \ref{tab:sar_buoy_validation}. The comparison results reveal that the Root Mean Square Errors (RMSE) for the six cases are 1.14 m/s, 0.98 m/s, 1.42 m/s, 1.17 m/s, 0.79 m/s, and 0.77 m/s, respectively, indicating that the retrieved wind speeds from the SAR images exhibit agreement with the buoy data. While SAR-derived wind speeds may not perfectly reproduce point-scale measurements, their accuracy and spatial consistency are sufficient for analyzing relative wind speed variations and wake-induced deficits over large offshore areas, which is the primary focus of this study.

\subsection{Upstream and downstream wind speed extraction}

\begin{figure*}[htb]
\centering
\includegraphics[width=16cm]{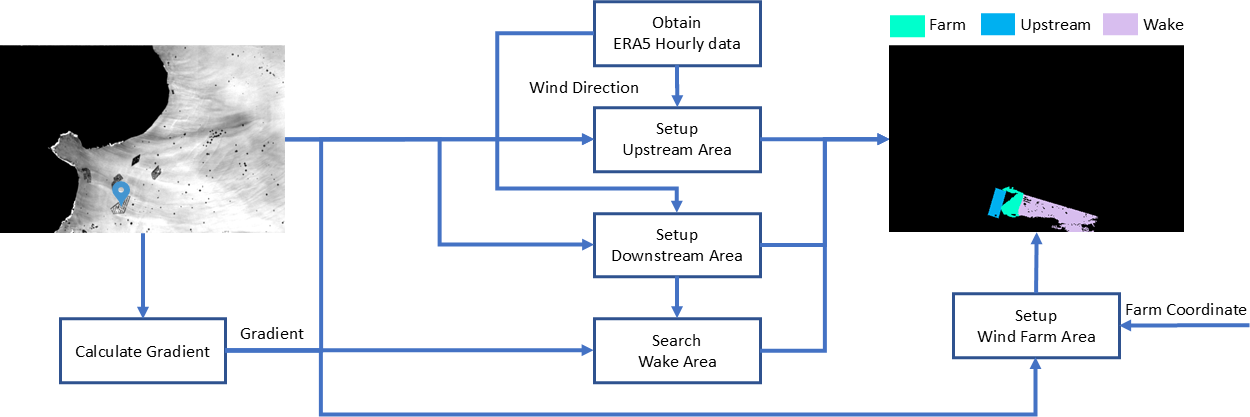}
\caption{\textbf{Workflow of the proposed \(R^2G^3\) for upstream and wake-affected areas detection.} Sentinel-1 SAR data are first used to calculate the spatial gradient of wind speed. ERA5 hourly data are employed to determine wind direction, which is then used to define upstream and downstream regions relative to the wind farm. Based on these regions, a search is conducted to identify wake-affected areas using gradient information. The wind farm area is delineated using known farm coordinates. The method integrates gradient-based analysis and geometric constraints to detect wake-affected regions. Colored regions represent the wind farm (green), upstream (blue), and wake-affected (purple) areas. Credit of the source SAR image: European Space Agency (ESA).}
\label{fig:4}
\end{figure*}

In this study, we develop an image processing method called Restricted-Region Gradient-Guided Growing (\(R^2G^3\)) to distinguish between the upstream free speed and the downstream wake speed from the retrieved wind speed data. During the pre-processing procedure, bright objects such as turbines and ships are masked out, which results in small or zero wind speed values in those areas after the multi-looking procedure, leading to abnormally high gradients. Similarly, areas affected by the wake effect typically have lower wind speeds compared to their surroundings, resulting in higher gradients at the boundary between wake and non-wake areas. Based on the above intuition, we calculate the gradient of the retrieved wind speed image to guide the identification of the Region of Interest (ROI) corresponding to the upstream and downstream areas, as shown in Fig. \ref{fig:4}. Obviously, the real ROIs are the upstream and downstream areas, excluding the wind farm itself. To achieve this, we use the location of the wind farm as the initial point and identify adjacent pixels with abnormal gradients as the farm area. Then, the local wind direction is extracted from the ERA5 hourly data \cite{hersbach2018era5}, which helps determine the orientations of the upstream and downstream areas. We assume that the upstream and downstream areas are rectangular in shape, with heights of 20 pixels (approximately 10 km) and 100 pixels (approximately 50 km), respectively. The widths of the rectangles are manually set according to the scale of the corresponding wind farm. For the upstream area, we filter out pixels with abnormally high gradients or abnormally small wind speeds, and the remaining pixels are considered the final upstream area. For the downstream area, we start with a few pixels located downstream of the wind farm (without abnormal gradients) as initial seeds. Based on these seeds, we iteratively expand the downstream area by including pixels whose wind speed values are close to the average speed of the seeds. The average speed is updated with the inclusion of new pixels. Finally, we determine the wake area after checking all pixels in the predefined downstream search area. By using the \(R^2G^3\) method, we can separate the upstream and downstream areas, allowing us to analyze and quantify the impact of near-surface wind farm wakes. The detailed procedure for detecting upstream and wake-affected areas is summarized in Algorithm \ref{alg1} with default settings reported in Table \ref{tab:R2G3_parameters}, while the algorithm overview and sensitivity test of parameters are provided in Supplementary Note 1 and Supplementary Note 2.

\begin{figure*}[htb]
\centering
\includegraphics[width=13cm]{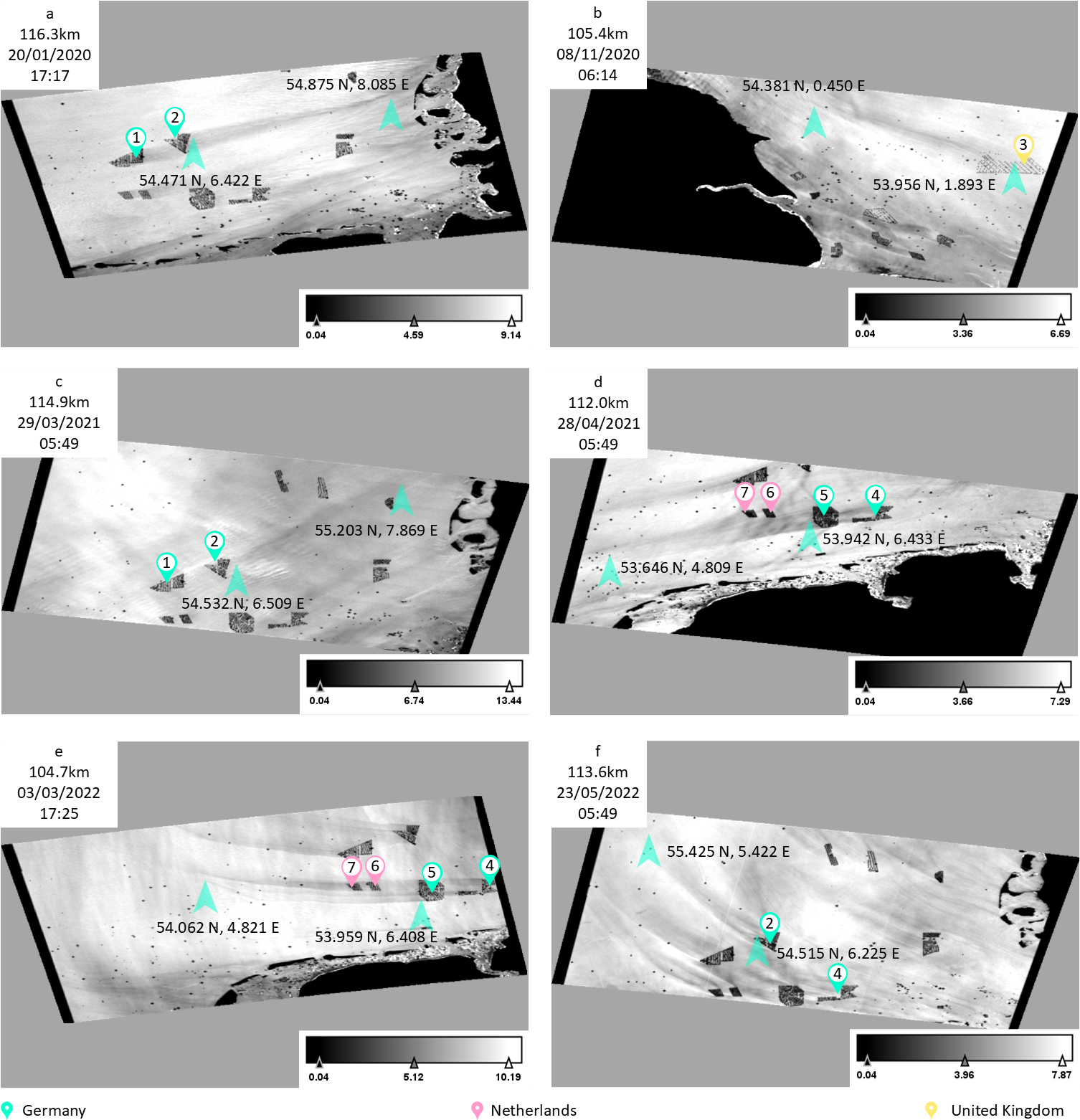}
\caption{\textbf{Examples of long-range near-surface wake signatures observed in SAR-derived 10 m wind speed fields.} The grayscale color bars represent SAR-retrieved neutral wind speed at 10 m height (m/s), with darker tones indicating lower wind speeds and lighter tones indicating higher wind speeds. Blue arrows mark the manually identified start and end points used to estimate the wake propagation distance, which is calculated based on geographic coordinates. The acquisition date and time (UTC) are shown in each subpanel. The SAR scenes were captured: (a) and (c)-(f) in the German Bight; and (b) near the coast of Holderness and Yorkshire. The color of the wind farm location symbols represents the operating country (see legend at bottom). Detailed information on the wind farms is provided in Table \ref{table:2}. Credit of the source SAR images: European Space Agency (ESA).}
\label{fig:5}
\end{figure*}

\begin{algorithm} 
	\caption{\(R^2G^3\) for upstream and wake-affected areas detection} 
	\label{alg1} 
	\begin{algorithmic}[1]
        \REQUIRE Wind Farm Coordinate $\mathbf{C}$, Wind Direction $D$, Wind Farm Radius $R$, Upstream Length $L_{\text{u}}$, Downstream Length $L_{\text{d}}$, Wake Speed Threshold $T_1, T_2$, Upstream Speed Threshold $T_{\text{u}}$
        
        \STATE Calculate the gradient $G$ of the wind speed field retrieved from the SAR image and set the threshold $G_{\text{T}}$
        
        \STATE Search pixels affected by wind turbines around $\mathbf{C}$ and exclude them from the subsequent analysis 
        
        \STATE Setup upstream area $A_{\text{u}}$ based on $D$, $R$ and $L_{\text{u}}$ and initialize upstream point list $P_{\text{u}} \gets \{\}$ 
        
        \STATE Setup downstream area $A_{\text{d}}$ based on $D$, $R$ and $L_{\text{d}}$
        
        \STATE Initialize the average wake wind speed $S_{\text{ave}}$ based on a point set $P_{\text{d}} = \{P_1, P_2, ..., P_N\}$ within the $A_{\text{d}}$
        
		\FOR{$i$ in $A_{\text{d}}$}
            \IF{$T_1 \times S_{\text{ave}} \leq S_i \leq T_2 \times S_{\text{ave}}$ and $G_i \leq G_{\text{T}}$}
                \STATE $P_{\text{d}} \gets \{P_1, ..., P_N, P_i\}$

                \STATE $S_{\text{ave}} \gets \text{Ave}(P_{\text{d}})$
            \ENDIF
        \ENDFOR
        
		\FOR{$j$ in $A_{\text{u}}$}
            \IF{$G_j \leq G_{\text{T}}$ and $S_j \geq T_{\text{u}} \times S_{\text{ave}}$}
                \STATE $P_{\text{u}} \gets P_{\text{u}} \cup \{P_j\}$
            \ENDIF
        \ENDFOR
        
        \STATE The final upstream area $P_{\text{u}}$ and wake-affected area $P_{\text{d}}$ are obtained.
	\end{algorithmic} 
\end{algorithm}

\section{Results}

\begin{table*}[t]
\centering
\caption{Multi-year verification of Sentinel-1 10-m wind speeds against in-situ buoy observations. Reported metrics include number of collocations ($N$), root-mean-square error (RMSE), mean bias (SAR--buoy), mean absolute error (MAE), Pearson correlation coefficient ($R$), scatter index (SI), and bootstrapped 95\% confidence intervals (CI) for RMSE and bias.}
\label{tab:sar_buoy_validation}
\begin{tabular}{lcccccccc}
\hline
Buoy & $N$ & RMSE & Bias & MAE & $R$ & SI & RMSE 95\% CI & Bias 95\% CI \\
     &     & (m s$^{-1}$) & (m s$^{-1}$) & (m s$^{-1}$) &  &  & (m s$^{-1}$) & (m s$^{-1}$) \\
\hline
Buoy1 & 117 & 1.14 & $-0.73$ & 0.93 & 0.95 & 0.15 & [1.00, 1.29] & [$-0.89$, $-0.56$] \\
Buoy2 & 304 & 0.98 & $-0.24$ & 0.68 & 0.92 & 0.20 & [0.81, 1.16] & [$-0.34$, $-0.13$] \\
Buoy3 & 46  & 1.42 & $-0.97$ & 1.10 & 0.95 & 0.18 & [1.11, 1.70] & [$-1.27$, $-0.68$] \\
Buoy4 & 84  & 1.17 & $-0.57$ & 0.93 & 0.95 & 0.15 & [0.97, 1.35] & [$-0.79$, $-0.36$] \\
Buoy5 & 77  & 0.79 & +0.07  & 0.59 & 0.96 & 0.14 & [0.62, 0.93] & [$-0.11$, $+0.25$] \\
Buoy6 & 317 & 0.77 & $-0.12$ & 0.56 & 0.94 & 0.13 & [0.68, 0.85] & [$-0.20$, $-0.03$] \\
\hline
\end{tabular}
\end{table*}

All analyses are based on SAR-retrieved 10 m neutral wind speeds. The results, therefore, describe near-surface wake signatures and should not be directly interpreted as hub-height wind conditions. Based on the analysis of SAR images obtained from over 60 wind farms across Europe and Asia during the years 2020, 2021, and 2022, two findings regarding wind farm wake effects are presented in this study. Firstly, the study demonstrates that near-surface wake effects can extend more than 100 km. The extended wake range has important implications for wind farm planning and development. Secondly, the study reveals that on average, wind farm wake effects lead to a 12.4\% reduction in 10 m wind speed in downstream wake-affected sea areas. While this reduction is quantified at the near-surface level, it may have implications for the power generation potential of downstream wind farms. The physical plausibility of the observed long-range and transboundary wake propagation is further supported by mesoscale simulations, as described in Supplementary Note 3.

\subsection{Long-range near-surface wake signatures confirmed by SAR}

Early research utilizing SAR data has provided evidence that wind farm wakes can persist for more than 20 km downstream \cite{christiansen2005wake}. Mesoscale studies using the Weather Research and Forecasting (WRF) model have also shown velocity deficits in downstream areas at distances of at least 15 km from upstream wind farms \cite{jimenez2015mesoscale}. However, with the increase in turbine size and farm scale, there is mounting evidence that estimates of 15 km or 20 km for wake propagation are no longer valid. In-situ measurements conducted over wind farms in the German Bight, such as Amrumbank West and Godewind, confirm that far wake effects can extend at least 45 km and can reach up to 70 km downwind \cite{platis2018first}. Qualitative studies based on SAR images have also revealed that wakes can extend over 50 km \cite{hasager2015using, canadillas2020offshore}. In the case of superimposed wakes behind multiple wind farms, the wake length can reach distances of up to 70 km, as observed in SAR images \cite{djath2018impact}. While previous SAR-, LiDAR-, and model-based studies have reported long wind farm wakes under specific conditions, these investigations were largely limited to individual cases or regional domains. Comprehensive observational evidence documenting the spatial extent of near-surface wake signatures across multiple offshore wind farm regions has remained limited.

Currently, the recovery of low momentum and high turbulence caused by upstream wind farms is deemed to require a distance of at least 50 km \cite{lundquist2019costs}. However, our study, based on SAR-retrieved wind speed, reveals observational evidence of long-range near-surface wakes extending up to about 100 km downstream under favorable atmospheric conditions, as shown in Fig. \ref{fig:5}. These long-distance near-surface wakes were captured in the North Sea at various times over a three-year period. ERA5-derived bulk Richardson numbers were analyzed for these six wake cases. All events occurred under very stable boundary-layer conditions, indicating strongly suppressed turbulent mixing. This finding is consistent with the observed long wake persistence \cite{platis2018first}, as reduced vertical momentum exchange delays wake recovery. The details of wind farms involved can be found in Table \ref{table:2}. Beyond the absolute wake extent, the SAR observations reveal that long-range wake persistence is strongly associated with wind farm clustering and cumulative installed capacity. In particular, near-surface wakes extending beyond 100~km predominantly occur where multiple large-scale wind farms are aligned with the prevailing wind direction, leading to superposition of individual wake signatures and delayed momentum recovery. This behavior is consistent with recent mesoscale modeling studies suggesting that wake recovery length scales increase nonlinearly with farm size and turbine density \cite{pryor2021wind, borgers2023mesoscale}, and the present results provide observational evidence consistent with these modeling results at regional scales. Meanwhile, the observation that a single ultra-large cluster, such as the Hornsea Project in Fig. \ref{fig:5} (b), can generate near-surface wake signatures exceeding 100~km further suggests that wake recovery assumptions commonly used in offshore planning may no longer be valid for next-generation wind farms. As turbine rotors increase in size and farms expand to gigawatt scales, wake-induced flow modification increasingly resembles a mesoscale phenomenon rather than a localized engineering effect.

At the same time, a considerable number of large-scale wind farms are currently being constructed or planned within the North Sea. For example, in the vicinity of wind farm cluster 1 in Fig. \ref{fig:5}, several wind farms with a combined capacity of more than 14 GW are already under planning. At the same time, in the vicinity of cluster 2, multiple German wind farms with a combined capacity of nearly 3 GW are currently under preliminary assessment. As shown in Fig. \ref{fig:5} (a) and Fig. \ref{fig:5} (c), the superimposed wakes caused by clusters 1 and 2 are already clearly visible and extend over a considerable distance. Upon completion of all the planned wind farms, with a capacity that is approximately nine times larger than the current capacity, the wake effects will cover a much wider geographical area and extend over a much greater distance. This substantial impact can have immediate economic repercussions and potentially hinder the development of future wind projects, reducing the sustainability of wind energy utilization. It becomes crucial to carefully consider and mitigate the potential challenges posed by these extended wake effects to ensure the sustainable growth and success of wind energy projects.

\subsection{Cross-boundary near-surface wake interactions revealed by SAR}

\begin{table*}[t]
\centering
\caption{Default parameters used in the \(R^2G^3\) wake extraction algorithm. All parameters are applied uniformly to all SAR scenes unless otherwise stated.}
\label{tab:R2G3_parameters}
\begin{tabular}{llll}
\hline
\textbf{Symbol} & \textbf{Description} & \textbf{Default value} & \textbf{Notes} \\
\hline

$\mathbf{C}$ & Wind farm center coordinate & Farm centroid & Derived from turbine layout \\

$R$ & Wind farm radius & Manually measured & Max distance from $\mathbf{C}$ to turbines \\

$D$ & Wind direction & ERA5 10\,m wind & Nearest time and location to SAR overpass \\

$L_{\text{u}}$ & Upstream length & 10 km & Along-wind direction \\
$L_{\text{d}}$ & Downstream length & 50 km & Along-wind direction \\
$W$ & Upstream/downstream width & Set based on $R$ & Cross-wind extent \\

$T_1$ & Lower wake speed ratio & 0.6 & $S_i \ge T_1 \times S_{\text{ave}}$ \\
$T_2$ & Upper wake speed ratio & 1.1 & $S_i \le T_2 \times S_{\text{ave}}$ \\
$T_{\text{u}}$ & Minimum upstream speed & $0.5 \times S_{\text{ave}} $ & Calculated based on the average waked wind speed \\

$G_{\text{T}}$ & Gradient threshold & 2 $\times$ average gradient & Derived from pixels with gradient range (0.1, 50) \\

$N$ & Initial downstream seed size & 25 pixels & Center point + 3 offset layers $\{1, 3, 5\}$ in 8 directions \\
$S_{\min}$ & Minimum valid wind speed & 1 m/s & Avoids low-wind SAR uncertainty \\
$S_{\max}$ & Maximum valid wind speed & 25 m/s & Avoids high-wind SAR uncertainty \\
$N_{\text{u,min}}$ & Minimum upstream area valid pixels & $5 \times R $ & Applied to $P_{\text{u}}$ \\
$N_{\text{d,min}}$ & Minimum wake area valid pixels & $20 \times R $ & Applied to $P_{\text{d}}$ \\
\hline
\end{tabular}
\end{table*}

In addition to the long propagation distance of wind farm wakes, the velocity deficits caused by wind farm wakes are also substantial. For onshore wind farms, an estimated power generation loss of 5\% has been observed, based on an econometric model and numerical weather prediction simulations \cite{lundquist2019costs}. Simulation studies for offshore wind farms have reported velocity deficits of 7\% extending 100 km downstream and 10\% extending 80 km downstream \cite{stoelinga2022estimating}. Real-world observations have shown persistent wake effects with velocity deficits of approximately 21\% at a distance of 55 km downstream, while the speed loss can reach as high as 41\% approximately 24 km downstream, as measured by LiDAR \cite{schneemann2020cluster}. Wake effects can even be observed at 200 m altitude, albeit with smaller deficits compared to the hub height measurements \cite{canadillas2022offshore}.

\begin{figure*}[htb]
\centering
\includegraphics[width=13cm]{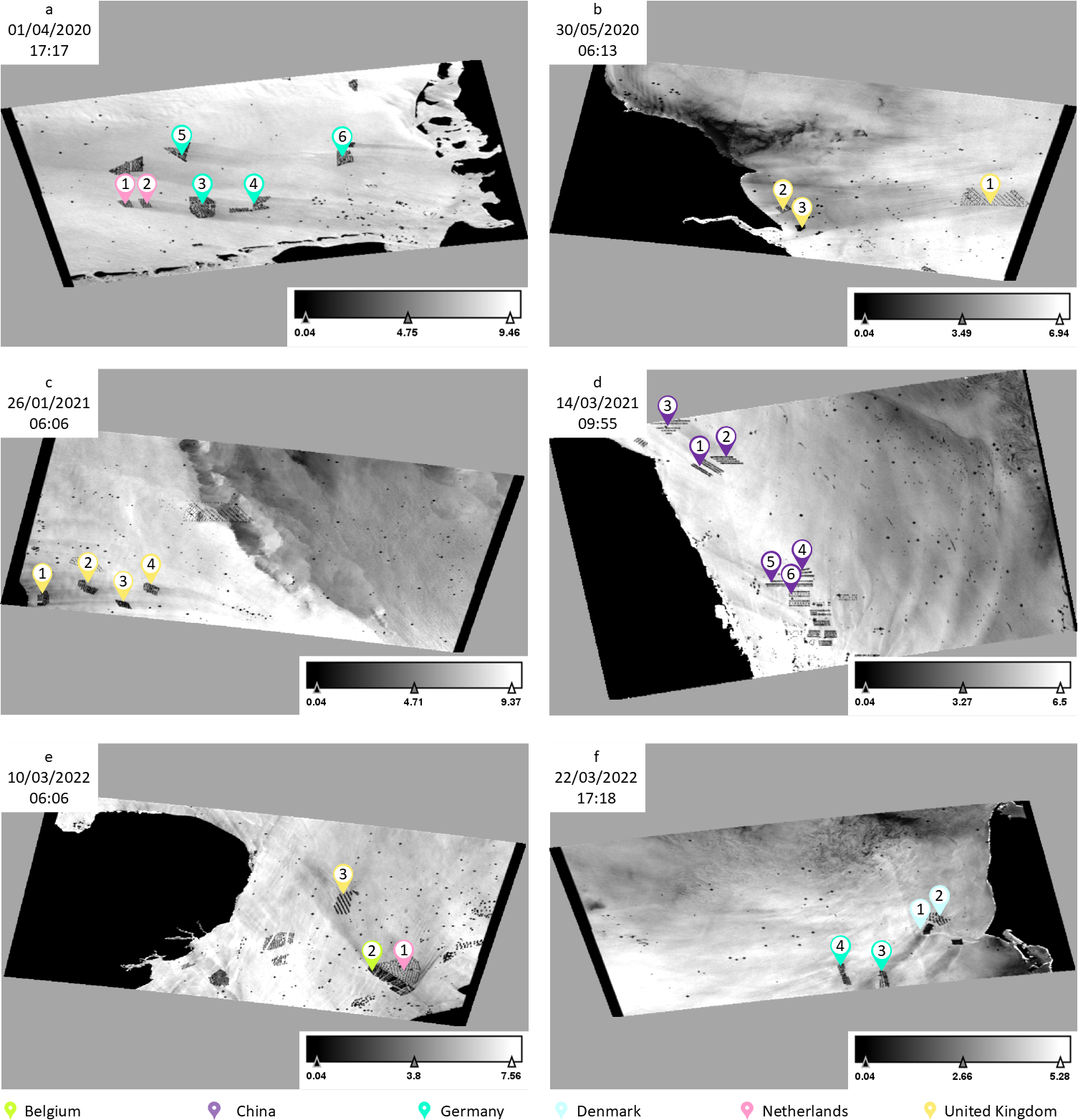}
\caption{\textbf{Examples of upstream wake impacts on downstream wind farms observed in SAR-derived 10 m wind speed fields.} The grayscale color bars represent SAR-retrieved neutral wind speed at 10 m height (m/s), with darker tones indicating lower wind speeds and lighter tones indicating higher wind speeds. The acquisition date and time (UTC) are shown in the top-left corner of each subpanel. The color of the wind farm location symbols corresponds to the operating country (see legend at bottom). The SAR scenes were captured: (a) in the German Bight; (b) near the coast of Holderness and Yorkshire; (c) near the coast of Lincolnshire; (d) in the Yellow Sea near Jiangsu Province; (e) in the North Sea between the UK and the Netherlands; and (f) near the Danish Jutland coast. Detailed information on the wind farms involved is provided in Table~\ref{table:3}. Credit of the source SAR images: European Space Agency (ESA).}
\label{fig:6}
\end{figure*}

Despite the extensive recognition and systematic investigation of wind farm wakes, the issue of justifying and quantifying their effects to resolve disputes between upstream and downstream wind farms remains unresolved \cite{finseraas2022gone, duvivier2020moat}. The absence of solid evidence to demonstrate the influence of wake effects is an urgent issue that needs to be addressed. While simulation results can suggest that wake effects may decrease productivity under certain conditions, they cannot prove the actual impacts on downstream wind farms in real-world operating scenarios, not to mention when and how these impacts occur. In contrast, SAR images offer a valuable and generic solution, as they capture wind speeds under real-world operating conditions, providing strong observational evidence of the actual near-surface impacts of upstream wakes. 

Fig. \ref{fig:6} visually represents the near-surface wake interactions between adjacent wind farms. Panels (b), (c), and (d) demonstrate the wake interactions among wind farms within the same country, typically owned by different companies. In panel (b), the downstream wind farm clusters are entirely enveloped by the long-range and high-intensity near-surface wake effects generated by the upstream large-scale Hornsea Project. Medium-scale wind farms located in close proximity, as shown in panels (c) and (d), also experience substantial near-surface wake interactions with each other. Resolving disputes between domestic wind farms can be relatively straightforward with the involvement of local authorities. However, complexities arise when wake interactions occur between wind farms owned by different countries, as evident in panels (a), (e), and (f). Taking panel (e) as an example, although three wind farm clusters are situated in their respective exclusive economic zones, the near-surface wake generated by clusters 1 (Netherlands) and 2 (Belgium) clearly affects downstream cluster 3 (United Kingdom). Table \ref{table:3} provides detailed information about the wind farms involved in wake effects, including those generating wakes and those influenced by wakes. It is noteworthy that a wind farm cluster can be both a culprit and a victim depending on the wind direction. Resolving transnational wake interactions requires bilateral and multilateral negotiation and cooperation to mitigate potential conflicts. From a scientific perspective, these transnational wake interactions also highlight a limitation of site-specific wake models that neglect upstream developments beyond national boundaries. The SAR observations demonstrate that near-surface wake impacts can propagate across exclusive economic zones, implying that national-scale planning frameworks may systematically underestimate external wake exposure in densely developed sea basins such as the North Sea. 

\subsection{A first multi-regional census on wake effects}

\begin{figure*}[!htb]
\centering
\includegraphics[width=16cm]{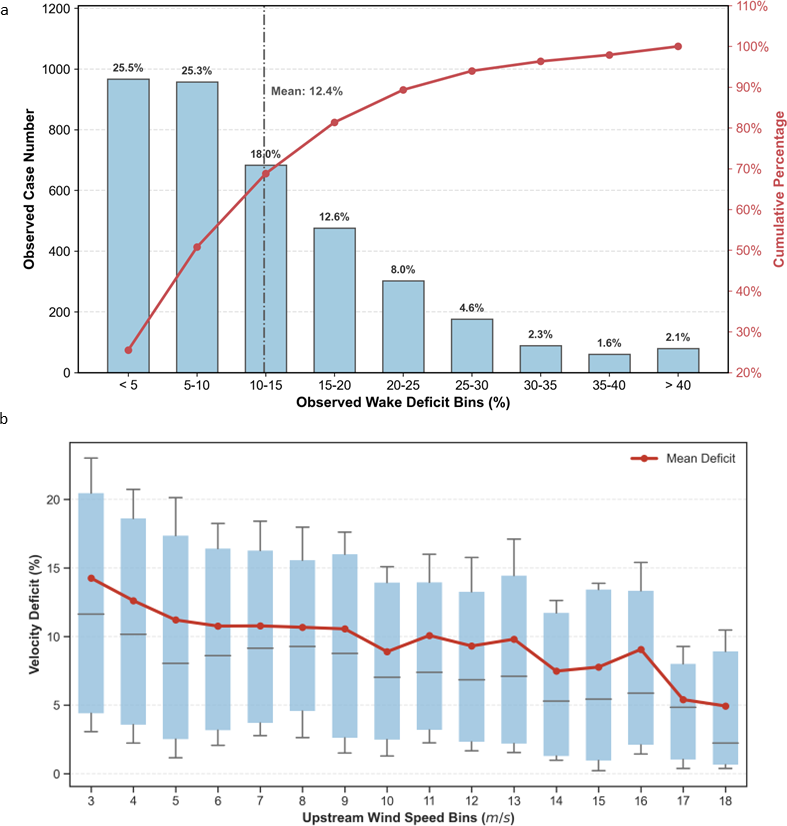}
\caption{\textbf{Distribution and characteristics of observed near-surface wind speed deficits.} (a) Distribution of observed near-surface wind speed deficits grouped by percentage intervals. Bars represent sample counts and the red curve denotes the cumulative percentage. (b) Observed velocity deficit as a function of upstream wind speed. Box plots show the inter-quartile range (25th–75th percentiles), with whiskers extending to the 20th and 80th percentiles. The red line indicates the mean deficit within each wind speed bin. n = 3789 samples.}
\label{fig:7}
\end{figure*}

After identifying wake-affected regions within individual SAR scenes, this section focuses on conducting a large-scale statistical assessment to quantify the associated near-surface wind speed deficits.

By utilizing the retrieved wind speed data and identifying the upstream and wake-affected areas, this study accomplishes the first multi-regional census on wind farm wake effects. Initially, cases with very small detected areas (often due to farms located at the edge of the SAR image) and extreme wind speeds (above 25 m/s or below 1 m/s) are filtered out, resulting in 5920 samples for the subsequent quantitative assessment. Among all these samples, 3789 (64.0\%) exhibit lower downstream wind speeds compared to the upstream wind speeds, indicating the presence of a near-surface wind speed deficit. As SAR is a real-world observation, there exist multiple potential factors that lead to the anomaly that the downstream wind speed is even higher than the upstream, such as the imaging quality, noise effect, side wind, retrieval quality and detection accuracy. And sometimes, this kind of phenomenon may indeed exist in the real world under certain atmospheric conditions \cite{hasager2023wind, larsen2021case, vanderwende2016simulating, yang2014large}, which needs further in-depth investigation. As the real reasons are unclear, cases where the downstream wind speed exceeds the upstream wind speed are excluded from the statistical analysis to reduce ambiguity. This filtering step may influence both the distribution and magnitude of the reported wind speed deficits and should therefore be considered when interpreting the statistical results. 

The average near-surface velocity deficit of the remaining samples is 0.990 m/s, which corresponds to a 12.4\% reduction compared to the upstream wind speeds. Fig. \ref{fig:7}(a) demonstrates that over half of the observed near-surface wind speed deficit are below 10\%, while approximately 10\% of the samples experience reductions exceeding 30\%. Furthermore, Fig. \ref{fig:7}(b) reveals a correlation between lower incoming wind speeds and increased wake-induced reductions, consistent with in-situ measurement results \cite{barthelmie2007modelling} and may be related to the aerodynamic operating regime of wind turbines. Typically, wind turbines operate with a higher thrust coefficient at wind speeds below the rated value to maximize power extraction, which tends to generate more intense wakes. In contrast, at above-rated wind speeds, pitch control mechanisms reduce the thrust coefficient, potentially leading to a relative weakening of the wake intensity. The observed dependence of wake-induced wind speed deficit on upstream wind speed has important implications for long-term energy yield assessments. Because lower wind speed conditions are more susceptible to wake losses \cite{owda2022wind}, cumulative wake impacts may disproportionately affect annual energy production during moderate and low-wind regimes that contribute substantially to total operating hours. This nonlinear interaction suggests that average wake losses inferred from idealized or high-wind conditions may underestimate real-world, long-term impacts in densely populated wind farm regions.

\begin{figure*}[htb]
\centering
\includegraphics[width=18cm]{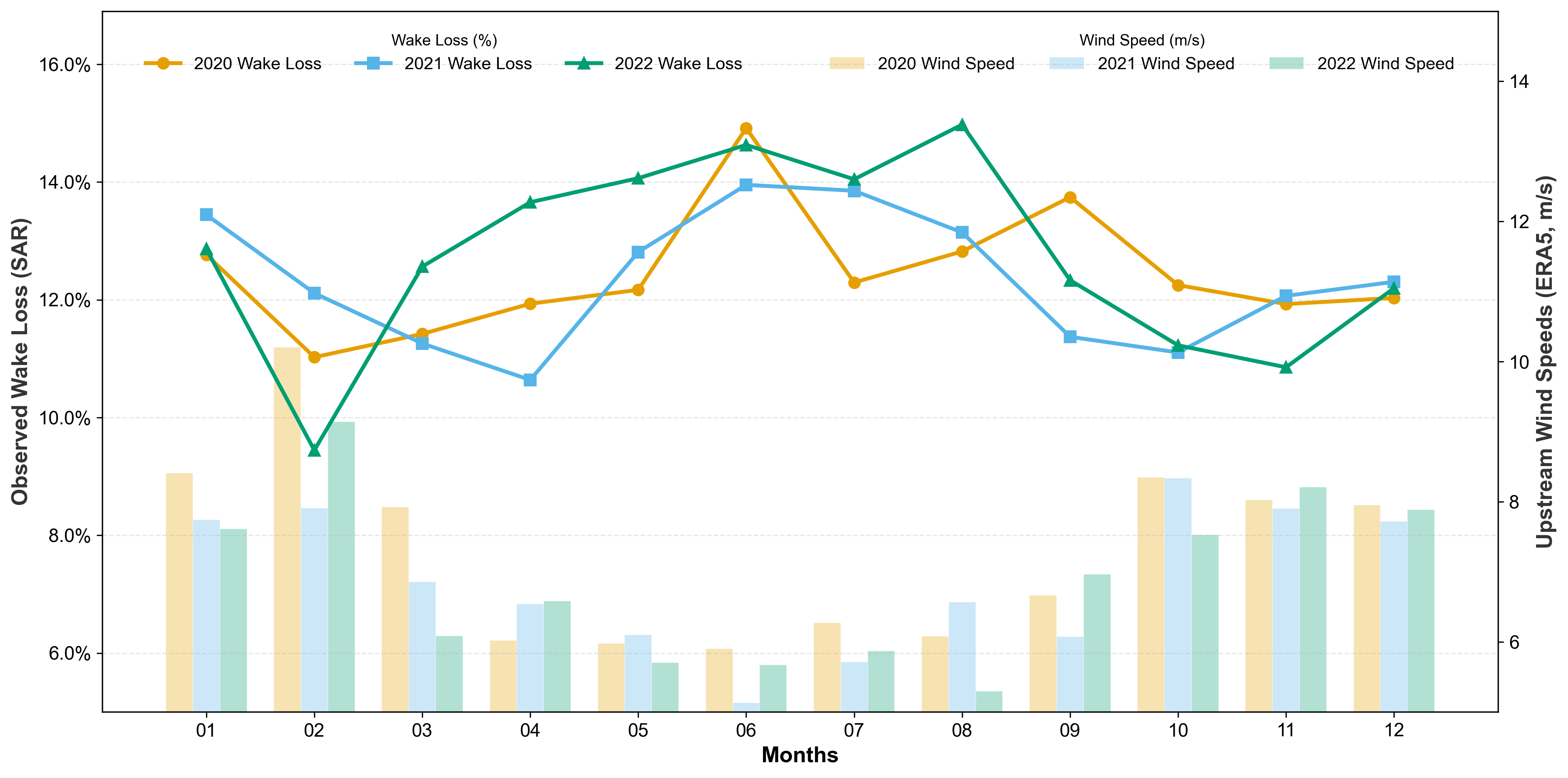}
\caption{\textbf{Monthly average wake-induced wind speed deficit.} The horizontal axis represents the month. The line indicates the near-surface wind speed deficit observed by SAR, while the bars represent the corresponding monthly mean wind speed derived from ERA5.}
\label{fig:8}
\end{figure*}

\begin{table*}[htb]
\setlength{\abovecaptionskip}{0.cm}
\caption{Information about wind farms contained in Fig. \ref{fig:5}, including operating country, capacity (MW) and the number of turbines. In total, 18 wind farms belonging to three countries (Germany, United Kingdom and Netherlands) are included, with 1269 turbines and 7.80 GW capacity.}
\label{table:2}
\centering
\begin{tabular}{ccccc}
\hline
Number             & Wind Farm Name            & Country             & Capacity (MW)        & Turbines             \\ \hline
\multirow{3}{*}{1} & Deutsche Bucht            & Germany             & 252                  & 31                   \\
                   & Veja Mate                 & Germany             & 402                  & 67                   \\
                   & BARD Offshore 1           & Germany             & 400                  & 80                   \\\hdashline
\multirow{3}{*}{2} & Hohe See                  & Germany             & 497                  & 71                   \\
                   & Albatros                  & Germany             & 112                  & 16                   \\
                   & Global Tech I             & Germany             & 400                  & 80                   \\\hdashline
\multirow{2}{*}{3} & Hornsea Project One       & United Kingdom      & 1218                 & 174                  \\
                   & Hornsea Project Two       & United Kingdom      & 1386                 & 165                  \\\hdashline
\multirow{3}{*}{4} & Nordsee One               & Germany             & 332.1                & 54                   \\
                   & Gode Wind 1               & Germany             & 330                  & 55                   \\
                   & Gode Wind 2               & Germany             & 252                  & 42                   \\\hdashline
\multirow{6}{*}{5} & Trianel Windpark Borkum 1 & Germany            & 200                  & 40                   \\
                   & Trianel Windpark Borkum 2 & Germany            & 203                  & 32                   \\
                   & Borkum Riffgrund 1        & Germany             & 312                  & 78                   \\
                   & Borkum Riffgrund 2        & Germany             & 450                  & 56                   \\
                   & Merkur                    & Germany             & 396                  & 66                   \\
                   & Alpha Ventus              & Germany             & 60                   & 12                   \\\hdashline
6                  & \multirow{2}{*}{Gemini}   & \multirow{2}{*}{Netherlands} & \multirow{2}{*}{600} & \multirow{2}{*}{150} \\
7                  &                           &                     &                      &                      \\ \hline
\end{tabular}
\end{table*}

Next, the monthly observed near-surface wake-induced wind speed deficits during the three years are presented in Fig.~\ref{fig:8}. The line graph shows that larger deficits are observed more frequently during the summer months compared to the winter months, which is consistent with previous studies \cite{djath2019wind}. To better interpret this seasonal pattern, we analyze the corresponding monthly mean background wind conditions using ERA5 hourly 10 m wind speed data \cite{hersbach2018era5}. It should be noted that ERA5 data are used only to characterize the seasonal background wind climate, while the wake-induced wind speed deficits themselves are derived entirely from SAR-retrieved 10 m wind speeds. The ERA5-derived monthly mean wind speeds over the study regions are shown as the bar graph in Fig.~\ref{fig:8}. The results indicate that the monthly mean wind speeds from April to September are generally lower than those from October to March. Considering the relationship between wind speed and wake deficit (Fig.~\ref{fig:7}(b)), where larger deficits tend to occur under lower ambient wind speeds, this seasonal variation in background wind conditions provides a plausible explanation for the more pronounced wake-induced deficits observed during the summer months.

\subsection{A first open-access SAR database for wind farm wakes}
The SAR images used in this study have been made publicly available. The dataset is organized according to acquisition time and wind farm location. In addition, a curated collection of representative inter-farm wake interactions identified from SAR imagery is provided to facilitate further research on this phenomenon by both academia and industry. We believe that such a database can serve as a foundation for researchers to further develop and benchmark SAR-based wake detection and assessment frameworks, as well as to validate wind farm wake parameterizations used in mesoscale and engineering models. The dataset may also support the development of machine-learning approaches for automated wake detection and classification, and enable multi-sensor studies that combine SAR observations with other remote sensing or in-situ measurements. Furthermore, integrating this dataset with mesoscale simulations could help expand temporal coverage and improve wind speed extrapolation techniques from near-surface observations to turbine-relevant heights. Overall, the database is expected to benefit the research community investigating long-range wake signatures and improve our understanding of wake interactions among large offshore wind farm clusters.

At the same time, it should be noted that all wake metrics in this study are derived from SAR-retrieved 10 m neutral wind speeds and therefore represent near-surface wake signatures. Wind-farm-induced wake behavior may vary with height, and wake characteristics in the surface layer may differ from those in the turbine rotor layer (about 50–150 m), particularly under stable atmospheric conditions with strong vertical shear. Previous studies combining SAR observations and mesoscale modeling have shown that near-surface wind speed deficits or speed-up may coexist with classical wake deficits at hub height, reflecting stability-dependent turbulence redistribution and vertical momentum exchange \cite{hasager2023wind}.

Consequently, long-range wake propagation observed at 10 m should not be interpreted as a direct proxy for turbine-level wake persistence or downstream power deficits. Instead, SAR-derived wake signatures provide valuable information on the large-scale, near-surface footprint of wind-farm-induced flow modification, which complements hub-height observations and modeling. Establishing quantitative relationships between near-surface and rotor-layer wake behavior remains an important topic for future research that requires coordinated satellite observations, in-situ measurements, and high-resolution numerical simulations.

\begin{figure*}[htb]
\centering
\includegraphics[width=16cm]{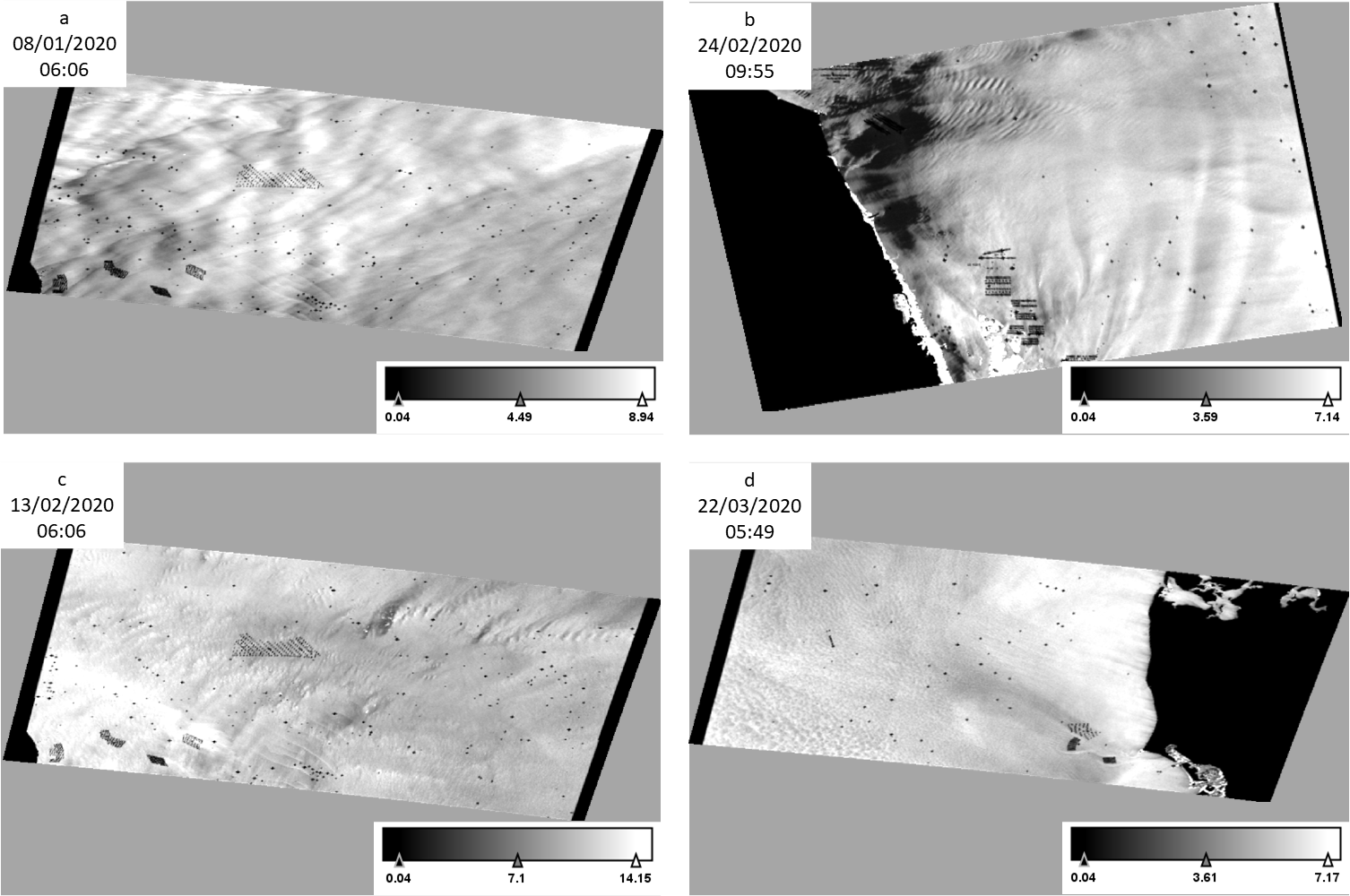}
\caption{\textbf{Examples of SAR-derived 10 m wind speed fields illustrating ambiguous boundaries between wake-affected and wake-free regions.} The grayscale color bars represent SAR-retrieved neutral wind speed at 10 m height (m/s), with darker tones indicating lower wind speeds and lighter tones indicating higher wind speeds. The SAR scenes were captured: (a) and (c) near the coast of Holderness and Yorkshire; (b) in the Yellow Sea near Jiangsu Province; and (d) near the Danish Jutland coast. The acquisition date and time (UTC) are shown in the top-left corner of each subpanel. Credit of the source SAR images: European Space Agency (ESA).}
\label{fig:9}
\end{figure*}

\subsection{Uncertainty in SAR-derived wake deficit estimates}

The CMOD5.N geophysical model function is known to exhibit different uncertainty characteristics depending on wind speed regime. At moderate wind speeds (approximately 4–15 m/s), CMOD5.N retrievals are generally robust, while increased uncertainty may occur under very low wind speeds due to reduced backscatter sensitivity and under high wind speeds due to signal saturation. In this study, extreme wind speed cases ($<1$ m/s and $>25$ m/s) were excluded from the quantitative analysis to reduce these effects. The remaining dataset predominantly lies within the well-validated operational range of CMOD5.N.

Table~\ref{tab:sar_buoy_validation} summarizes a multi-year validation of Sentinel-1 SAR 10-m wind speeds retrieved using CMOD5.N against in-situ buoy observations. In addition to RMSE and mean bias, several complementary metrics are reported, including mean absolute error (MAE), correlation coefficient ($R$), and scatter index (SI), providing a more complete characterization of retrieval performance. To quantify statistical robustness, 95\% confidence intervals for RMSE and bias were estimated using a bootstrap resampling approach. Across all buoy stations, RMSE values typically remain below approximately 1.0 m/s, with relatively narrow confidence intervals, indicating stable and consistent retrieval accuracy over different locations and sampling sizes. Correlation coefficients exceeding 0.9 further confirm the strong agreement between SAR-derived and in-situ wind speeds.

While Table~\ref{tab:sar_buoy_validation} characterizes the absolute uncertainty of SAR-retrieved wind speeds, wake deficits in this study are defined as the difference between spatially averaged upstream and downstream wind speeds extracted from the same SAR acquisition. As a result, scene-level systematic errors in SAR wind retrieval are expected to partially cancel when computing the differential wake signal. The remaining uncertainty is therefore dominated by random retrieval noise, which is further reduced through spatial averaging over hundreds of pixels within the \(R^2G^3\)-identified regions. Consequently, the effective uncertainty of wake deficit estimates will be smaller than the absolute wind speed retrieval error reported in Table~\ref{tab:sar_buoy_validation}. And importantly, the reported mean wake deficit should be interpreted as a climatological average across thousands of independent SAR observations, rather than a deterministic estimate for individual wind farm pairs or specific meteorological conditions.

\section{Discussion}

Accurately identifying wakes in real-world SAR imagery, such as those shown in Fig. \ref{fig:9}, remains a major challenge. Consequently, a thorough and precise assessment of long-range wake signatures demands further collective effort from the wider community. Looking ahead, several important directions are described below. 

First, all current investigations are based on wind speeds at a height of 10 m, whereas typical turbine hub heights exceed 100 m. At present, there are no suitable tools or algorithms for extrapolating wind speed from 10 m to 100 m. To overcome this challenge, it is important to collect in-situ wind speed measurements at 100 m in wake-affected regions and develop a practical algorithm to extend SAR-retrieved wind speeds to hub height.

Second, although the reduction in wind speed over distance caused by the wake is a key metric for wind farm planning, the irregular shape of the wake poses substantial measurement difficulties. Consequently, novel methods for measuring distance in this context will be an important area to enable SAR-supported wind farm planning.

Third, wind turbines operate under different performance regimes. For example, under above-rated wind conditions, power output is regulated and may remain approximately constant despite variations in wind speed, which can influence wake characteristics. Future research that incorporates turbine operational information could help better understand how different operating regimes affect wake intensity and downstream wind speed deficits.

Fourth, wake characteristics vary under different atmospheric stability conditions. While atmospheric stability indicators can be readily obtained from reanalysis products, systematically classifying and interpreting wake behavior under different stability regimes across thousands of SAR scenes remains methodologically challenging. Future studies combining stability metrics with automated wake characterization could provide a more comprehensive understanding of stability-dependent wake dynamics.

Finally, the proposed \(R^2G^3\) metric still relies on manually selected parameters. The reason is that in real-world SAR imagery, even experienced experts often struggle to delineate the boundary between wake-free and wake-affected areas as it is not always clear and recognizable such as the examples shown in Fig \ref{fig:9}. In addition, turbine operational variability (e.g., temporary curtailment or partial loading) may reduce wake contrast in some cases, although such information is not directly observable from SAR data. A potential solution is to employ high-resolution simulations to model the wind field in the same region with and without turbine parameterization. Comparing these two scenarios could help more objectively delineate wake-affected areas and improve the robustness of wake identification.

To facilitate the use of the data from this paper, we will provide as much relevant information and data as possible for the wider communities, including wake labels generated by \(R^2G^3\) as well as the parameters to generate the label, wake statistics for each SAR image, wind farm locations, turbine coordinates, SAR acquisition timestamps, and more. We envisage that these resources will lead to deeper insights into wake interactions among large-scale wind farms.

\section{Conclusions}

This study confirms that near-surface wake signatures at 10 m height, as observed by SAR, can persist over distances exceeding 100 km in real-world offshore wind farm clusters. As a result, it suggests that the wake-free distance to guide new farm planning in the wind energy industry may need to be reexamined and revised. It should be emphasized that all results are based on SAR-retrieved 10 m neutral wind speeds. Although these near-surface wake signatures provide robust observational evidence of long-range wake influence, they cannot be directly extrapolated to turbine hub height or rotor-layer flow without additional modeling or in-situ measurements. Furthermore, this research represents the first international study to investigate 3,789 cases using SAR-derived wind speed data. The study covers more than 60 wind farms across the globe and spans a three-year duration from 2020 to 2022. The results reveal that the average near-surface velocity deficit caused by wake effects is 0.990 m/s, equivalent to 12.4\% of the upstream wind speeds. The substantial evidence obtained in this work provides strong observational support for the frequent and pronounced near-surface impact of wind farm wake signatures. This is particularly pronounced in sea areas densely populated with wind farm, such as the North Sea and Yellow Sea. These findings are expected to accelerate the progress of potential legislation and conflict resolution regarding wake-induced revenue losses between neighboring wind farms in the foreseeable future.

\section*{Data availability}
The list of SAR images used in this study is available at \url{https://github.com/lironui/SAR_WAKE.git}. The SAR data are available at the Copernicus Open Access Hub, \url{https://browser.dataspace.copernicus.eu/}. The ERA5 data are available at the European Center for Medium-Range Weather Forecasts, \url{https://www.ecmwf.int/en/forecasts/dataset/ecmwf-reanalysis-v5}. The buoy data are available at the National Data Buoy Center, \url{https://www.ndbc.noaa.gov}. The SNAP toolbox is available at Science Toolbox Exploitation Platform, \url{http://step.esa.int/main}. The WRF model is available at \url{https://github.com/wrf-model/WRF/releases}.

\section*{Code availability}
The code for the proposed Restricted-Region Gradient-Guided Growing (\(R^2G^3\)) is available in the public repository at \url{https://github.com/lironui/SAR_WAKE.git}.

\section*{CRediT authorship contribution statement}
Rui Li proposed and refined the concept, independently designed and implemented the whole processing framework including the SAR image processing, wind speed retrieval and verification, and wake effects extraction, designed and implemented the dataset analysis and visualization, configured and conducted the WRF simulation, led the overall research workflow and wrote the paper draft. Jincheng Zhang contributed to rendering the initial concept, analyzing and interpreting the results, refining the wake extraction, and refining and editing the manuscript. Xiaowei Zhao contributed to acquiring funding, analyzing and interpreting the results, and refining and editing the manuscript. Xiaowei Zhao supervised the overall research.

\section*{Funding statement}
Rui Li, Jincheng Zhang and Xiaowei Zhao disclose support for the research of this work from the UK Engineering and Physical Sciences Research Council (grant number: EP/Y016297/1).

\section*{Acknowledgments}
We express great appreciation to Copernicus Open Access Hub for providing the Sentinel images, Copernicus Climate Change Service for providing the ERA5 data, the National Data Buoy Center for providing buoy measurement data, the European Space Agency for providing the SNAP toolbox and the National Center for Atmospheric Research for providing WRF model.

\section*{Ethics Declarations}
\noindent \textbf{Competing interests} \newline
The authors declare no competing interests.

\begin{table*}[htb]
\centering
\setlength{\abovecaptionskip}{0.cm}
\caption{Information about wind farms contained in Fig. \ref{fig:6}, including operating country, capacity (MW) and the number of turbines. In total, 52 wind farms related to wake effects (to influence or to be influenced) belonging to six countries (Belgium, China, Germany, Denmark, Netherlands and United Kingdom) are included, with 3004 turbines and 16.73 GW capacity.}
\label{table:3}
\begin{tabular}{cccccc}
\hline
Panel              & Number              & Wind Farm Name            & Country             & Capacity (MW)        & Turbines             \\ \hline
\multirow{17}{*}{a} & 1                   & \multirow{2}{*}{Gemini}   & \multirow{2}{*}{Netherlands} & \multirow{2}{*}{600} & \multirow{2}{*}{150} \\
                    & 2                   &                           &                     &                      &                      \\\cdashline{2-2}[1pt/1pt]
                    & \multirow{6}{*}{3}  & Trianel Windpark Borkum 1 & Germany             & 200                  & 40                   \\
                    &                     & Trianel Windpark Borkum 2 & Germany             & 203                  & 32                   \\
                    &                     & Borkum Riffgrund 1        & Germany             & 312                  & 78                   \\
                    &                     & Borkum Riffgrund 2        & Germany             & 450                  & 56                   \\
                    &                     & Merkur                    & Germany             & 396                  & 66                   \\
                    &                     & Alpha Ventus              & Germany             & 60                   & 12                   \\\cdashline{2-2}[1pt/1pt]
                    & \multirow{3}{*}{4}  & Nordsee One               & Germany             & 332.1                & 54                   \\
                    &                     & Gode Wind 1               & Germany             & 330                  & 55                   \\
                    &                     & Gode Wind 2               & Germany             & 252                  & 42                   \\\cdashline{2-2}[1pt/1pt]
                    & \multirow{3}{*}{5}  & Hohe See                  & Germany             & 497                  & 71                   \\
                    &                     & Albatros                  & Germany             & 112                  & 16                   \\
                    &                     & Global Tech I             & Germany             & 400                  & 80                   \\\cdashline{2-2}[1pt/1pt]
                    & \multirow{3}{*}{6}  & Meerwind Süd/Ost          & Germany             & 288                  & 80                   \\
                    &                     & Nordsee Ost               & Germany             & 295.2                & 48                   \\
                    &                     & Kaskasi                   & Germany             & 342                  & 38                   \\\hdashline
\multirow{4}{*}{b}  & \multirow{2}{*}{1}  & Hornsea Project One       & United Kingdom      & 1218                 & 174                  \\
                    &                     & Hornsea Project Two       & United Kingdom      & 1386                 & 165                  \\\cdashline{2-2}[1pt/1pt]
                    & 2                   & Westermost Rough          & United Kingdom      & 210                  & 35                   \\
                    & 3                   & Humber Gateway            & United Kingdom      & 219                  & 73                   \\\hdashline
\multirow{5}{*}{c}  & \multirow{2}{*}{1}  & Inner Dowsing             & United Kingdom      & 97.2                 & 27                   \\
                    &                     & Lincs                     & United Kingdom      & 270                  & 75                   \\\cdashline{2-2}[1pt/1pt]
                    & 2                   & Race Bank                 & United Kingdom      & 573.3                & 91                   \\
                    & 3                   & Sheringham Shoal          & United Kingdom      & 316.8                & 88                   \\
                    & 4                   & Dudgeon                   & United Kingdom      & 402                  & 67                   \\\hdashline
\multirow{6}{*}{d}  & 1                   & Datang Jiangsu Binhai     & China               & 300                  & 95                   \\
                    & 2                   & SPIC Binhai South H3      & China               & 300                  & 75                   \\
                    & 3                   & SPIC Binhai North H2      & China               & 400                  & 100                  \\
                    & 4                   & Huaneng Sheyang H1        & China               & 301.5                & 67                   \\
                    & 5                   & Longyuan Sheyang H2       & China               & 301.5                & 67                   \\
                    & 6                   & SPIC Jiangsu Dafeng H3    & China               & 302.4                & 72                   \\\hdashline
\multirow{17}{*}{e} & \multirow{5}{*}{1}  & Borssele 1                & Netherlands         & 376                  & 47                   \\
                    &                     & Borssele 2                & Netherlands         & 376                  & 47                   \\
                    &                     & Borssele 3                & Netherlands         & 351.5                & 37                   \\
                    &                     & Borssele 4                & Netherlands         & 380                  & 40                   \\
                    &                     & Borssele Site V           & Netherlands         & 19                   & 2                    \\\cdashline{2-2}[1pt/1pt]
                    & \multirow{11}{*}{2} & Seamade (Mermaid)         & Belgium             & 235                  & 28                   \\
                    &                     & Northwester 2             & Belgium             & 219                  & 23                   \\
                    &                     & Nobelwind                 & Belgium             & 165                  & 50                   \\
                    &                     & Belwind                   & Belgium             & 165                  & 55                   \\
                    &                     & Seamade (SeaStar)         & Belgium             & 252                  & 30                   \\
                    &                     & Northwind                 & Belgium             & 216                  & 72                   \\
                    &                     & Rentel                    & Belgium             & 309                  & 42                   \\
                    &                     & Norther                   & Belgium             & 369.6                & 44                   \\
                    &                     & Thornton Bank - phase I   & Belgium             & 30                   & 6                    \\
                    &                     & Thornton Bank - phase II  & Belgium             & 184.5                & 30                   \\
                    &                     & Thornton Bank - phase III & Belgium             & 110.7                & 18                   \\\cdashline{2-2}[1pt/1pt]
                    & 3                   & East Anglia ONE           & United Kingdom      & 714                  & 102                  \\\hdashline
\multirow{4}{*}{f}  & 1                   & Horns Rev 2               & Denmark             & 209.3                & 91                   \\
                    & 2                   & Horns Rev 3               & Denmark             & 406.7                & 49                   \\
                    & 3                   & DanTysk                   & Germany             & 288                  & 80                   \\
                    & 4                   & Sandbank                  & Germany             & 288                  & 72                   \\ \hline
\end{tabular}
\end{table*}

\clearpage

\begin{center}
    {\Large\bfseries Supplementary Information for\par}
    \vspace{0.5em}
    {\LARGE\bfseries
    Long-range near-surface wake signatures of offshore wind farm clusters
    revealed by satellite observations\par}
\end{center}

\vspace{2em}

\renewcommand{\thesection}{Supplementary Note \arabic{section}}

\section{\(R^2G^3\) Algorithm Overview}
The Restricted-Region Gradient-Guided Growing (\(R^2G^3\)) algorithm is designed to extract upstream reference regions and downstream wake-affected regions from SAR-retrieved wind speed fields in a consistent and scalable manner. The algorithm operates on a single SAR scene at a time and consists of four main steps: (i) wind farm localization and masking, (ii) upstream and downstream search region definition, (iii) gradient-constrained region growing for wake identification, and (iv) quality control and filtering.

Table 5 summarizes the default parameter values used in the \(R^2G^3\) algorithm. Unless otherwise stated, these parameters are applied uniformly to all SAR scenes analyzed in this study to ensure methodological consistency between wind farms, regions, and seasons. The wind farm radius $R$ is defined as the maximum distance from the farm centroid to the outermost turbine and is measured once for each wind farm based on turbine layout. This parameter is independent of SAR imagery and meteorological conditions and therefore does not introduce scene-specific tuning. Certain parameters in the \(R^2G^3\) algorithm are defined adaptively to account for scene-to-scene variability in SAR-derived wind fields. For example, the gradient threshold $G_T$ is computed as a multiple of the mean wind speed gradient estimated from pixels within a predefined gradient range (0.1–50), excluding turbine-contaminated areas. This adaptive definition allows the algorithm to adjust to variations in overall wind speed level and SAR noise while maintaining a consistent relative sensitivity across all scenes. 

Importantly, parameter values were selected based on empirical inspection of representative SAR scenes and were fixed prior to large-scale analysis to avoid scene-specific tuning. 

\section{\(R^2G^3\) Parameter Sensitivity}
\label{AppB}

To assess the robustness of the \(R^2G^3\) algorithm to user-defined parameter choices, a sensitivity analysis was conducted by systematically varying key parameters, including the upstream length, downstream length, and the lower and upper wake speed ratio thresholds, while keeping all other parameters fixed at their default values. For each configuration, wake deficits were recomputed across the full multi-year dataset, and the resulting mean wake speed ratio and corresponding 95\% confidence intervals were recorded.

For the downstream length, values ranging from 30 to 100~km were examined. The mean wake speed ratio shows a gradual and monotonic increase with increasing downstream length; however, variations remain small and the 95\% confidence intervals largely overlap across the tested range. This indicates that the algorithm is not highly sensitive to the precise choice of downstream length, provided it is sufficiently large to capture the undisturbed inflow region.

Similarly, the upstream length was varied between 5 and 15~km. The resulting mean wake speed ratios and confidence intervals show only minor differences, suggesting that the wake extraction is robust to reasonable changes in the upstream extent used for wake characterization.

The sensitivity to the lower wake speed ratio threshold was tested for values between 0.5 and 0.7. The results indicate stable performance for thresholds of 0.5 and 0.6, while a slightly lower mean value is obtained for 0.7, reflecting the exclusion of weaker wake signals. Nevertheless, the overall variations remain limited, demonstrating that the algorithm performance is not dominated by the exact choice of this threshold.

For the upper wake speed ratio threshold, a wider sensitivity is observed. A low upper threshold (1.05) leads to a pronounced increase in the mean wake deficit, as the stricter criterion tends to retain only pixels with stronger wind speed reductions. In contrast, thresholds of 1.10 and 1.15 yield consistent results, with overlapping confidence intervals. This supports the use of an upper threshold of 1.10 as a balanced choice between wake detection sensitivity and noise rejection.

Overall, the parameter sensitivity analysis demonstrates that the \(R^2G^3\) algorithm is robust to reasonable variations in its key parameters. The default parameter set adopted in this study lies within statistically stable regions of the parameter space, ensuring that the reported wake characteristics are not artifacts of specific parameter choices.

\section{Physical Consistency Assessment}
\label{AppC}

\renewcommand{\figurename}{Supplementary Figure}
\renewcommand{\thefigure}{\arabic{figure}}
\setcounter{figure}{0}
\begin{figure*}[htb]
\centering
\includegraphics[width=18cm]{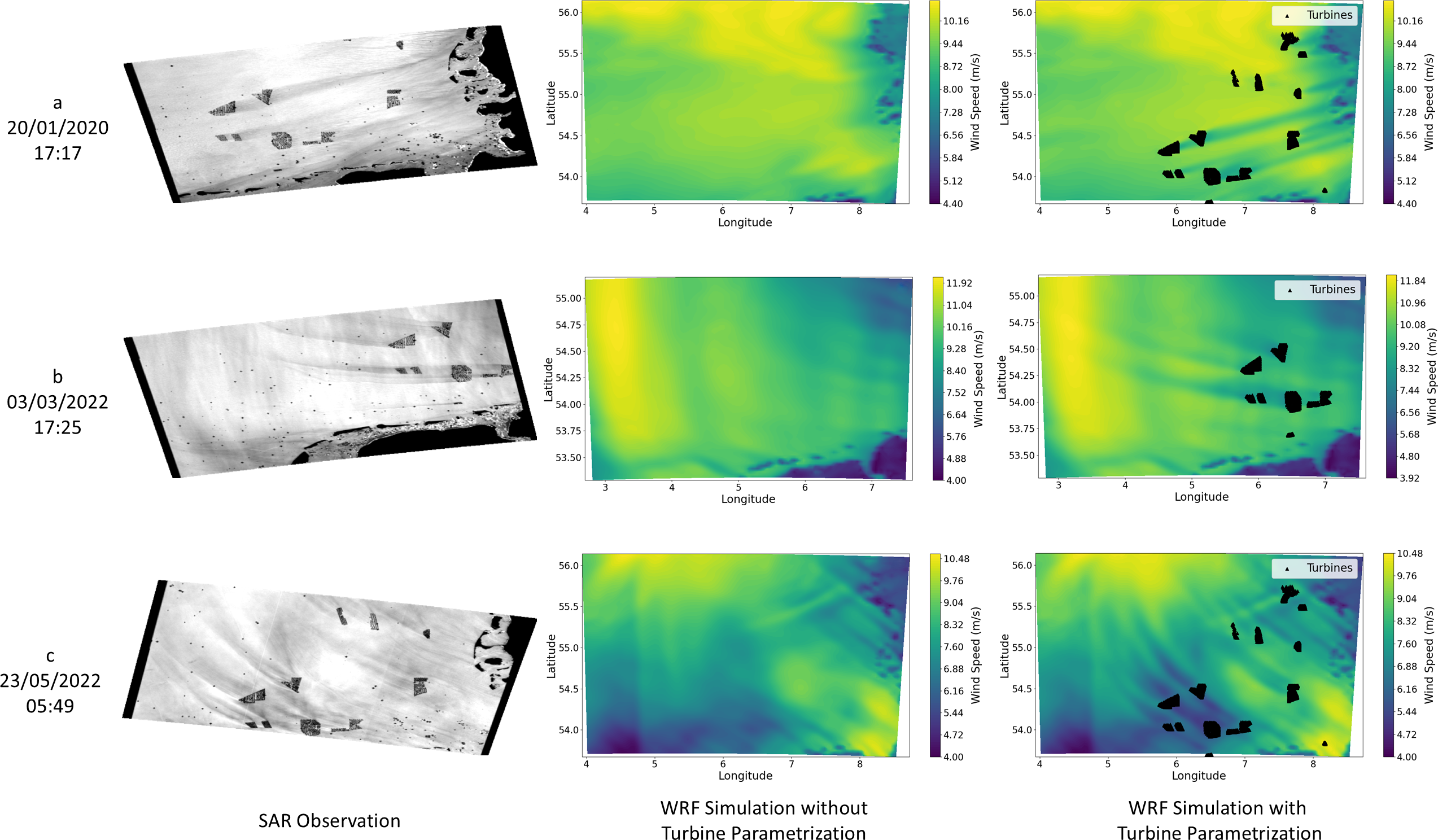}
\caption{\textbf{Validation of simulated long-range and transboundary wakes using SAR observations and WRF sensitivity simulations.} Rows (a–c) show three representative cases on 20 January 2020 (17:17), 3 March 2022 (17:25), and 23 May 2022 (05:49), respectively. The first column presents Sentinel-1 SAR observations. The second column shows WRF simulations without wind farm parameterization, and the third column shows WRF simulations with wind farm parameterization. Color shading represents wind speed, and black regions indicate wind turbine locations. The inclusion of wind farm parameterization in WRF enables the simulation of wake effects, which are compared with SAR-observed wake structures. Credit of the source SAR images: European Space Agency (ESA).}
\label{fig:10}
\end{figure*}

To assess the physical plausibility of the long-range and transboundary wake signatures identified in SAR imagery, mesoscale simulations were performed using the Weather Research and Forecasting (WRF) model (v4.6.1) \cite{skamarock2019description}. The objective is to evaluate whether a physical numerical framework reproduces the observed cross-boundary wake propagation under similar meteorological conditions.

WRF was configured with three nested domains at 27, 9, and 3 km horizontal resolution. The innermost 3 km domain covers the offshore wind farm clusters. We employed 62 vertical levels with enhanced resolution in the lower troposphere (the first level is about 10 m above ground level) to accurately resolve boundary-layer interactions and turbine-induced momentum extraction. Boundary layer processes were parameterized using the Mellor-Yamada-Nakanishi-Niino (MYNN) Level 2.5 scheme, which is well-suited for simulating turbine-induced turbulence. Other standard physics include the Thompson microphysics, the Rapid Radiative Transfer Model for General Circulation Models (RRTMG) radiation scheme, and the Noah Land Surface Model (LSM). Initial and boundary conditions were derived from ERA5 reanalysis data \cite{hersbach2018era5} at 3-hour intervals, with spectral nudging applied in the outer domain to maintain large-scale atmospheric fidelity. We implemented the Ma et al. (2022) scheme \cite{ma2022jensen}, which incorporates the XA (Xie-Archer) wake model \cite{xie2015self} and a sophisticated hub-height wind speed superposition method. This approach accounts for subgrid-scale wake interactions within farm clusters, providing an alternative representation of subgrid-scale wake interactions compared with conventional linear superposition approaches. Turbine specifications and locations were identical to those used in the SAR analysis. Two simulations were performed for each selected case: a control simulation without turbine parameterization and a wind farm simulation with turbine parameterization activated. All other model settings were kept identical between the two experiments, enabling direct comparison of wind fields with and without wind farm effects. 

As demonstrated by the contrast between the control and turbine-parameterized simulations in Fig.~\ref{fig:10}, elongated low-wind-speed structures emerge only when wind farm parameterization is activated. In the control simulations, no coherent downstream wake structures are evident at comparable scales. The spatial alignment between SAR-retrieved wake signatures and WRF-simulated wake structures provides independent dynamical support for the physical plausibility of the observed long-range and transboundary wake propagation. While differences in magnitude and fine-scale structure are expected due to model resolution (3 km), parameterization assumptions, and the inherent differences between grid-averaged model output and SAR-retrieved 10 m neutral winds, the consistency in propagation direction and spatial extent indicates that the satellite-observed wake features are unlikely to arise solely from satellite sampling geometry. These results suggest that as offshore wind farm clusters expand in scale and density, their wake influence may extend across jurisdictional boundaries under favorable atmospheric conditions. This highlights the potential relevance of coordinated maritime spatial planning and cross-border resource management, particularly in semi-enclosed basins such as the North Sea. 

While the present simulations demonstrate physical consistency for selected case studies, a comprehensive climatological assessment of transboundary wake frequency would require long-term multi-year simulations coupled with directional statistics, which we identify as an important direction for future work.

\renewcommand{\thetable}{\arabic{table}}
\setcounter{table}{0}
\begin{table}[t]
\centering
\caption{Sensitivity of the \(R^2G^3\) algorithm to the downstream length parameter.}
\label{tab:sensitivity_downstream}
\begin{tabular}{cccc}
\hline
\textbf{Downstream length} & \textbf{Mean} & \textbf{95\% CI (L)} & \textbf{95\% CI (H)} \\
\hline
30  & 0.902 & 0.874 & 0.930 \\
40  & 0.958 & 0.930 & 0.986 \\
50  & 0.990 & 0.961 & 1.018 \\
60  & 1.008 & 0.980 & 1.037 \\
70  & 1.020 & 0.991 & 1.049 \\
80  & 1.033 & 1.004 & 1.063 \\
90  & 1.041 & 1.011 & 1.070 \\
100 & 1.043 & 1.014 & 1.073 \\
\hline
\end{tabular}
\end{table}

\begin{table}[t]
\centering
\caption{Sensitivity of the \(R^2G^3\) algorithm to the upstream length parameter.}
\label{tab:sensitivity_upstream}
\begin{tabular}{cccc}
\hline
\textbf{Upstream length} & \textbf{Mean} & \textbf{95\% CI (L)} & \textbf{95\% CI (H)} \\
\hline
5  & 0.979 & 0.948 & 1.010 \\
10 & 0.990 & 0.961 & 1.018 \\
15 & 1.012 & 0.983 & 1.041 \\
\hline
\end{tabular}
\end{table}

\begin{table}[t]
\centering
\caption{Sensitivity of the \(R^2G^3\) algorithm to the lower wake speed ratio threshold.}
\label{tab:sensitivity_lower_ratio}
\begin{tabular}{cccc}
\hline
\textbf{Lower wake speed ratio} & \textbf{Mean} & \textbf{95\% CI (L)} & \textbf{95\% CI (H)} \\
\hline
0.5 & 0.990 & 0.962 & 1.018 \\
0.6 & 0.990 & 0.961 & 1.018 \\
0.7 & 0.956 & 0.928 & 0.983 \\
\hline
\end{tabular}
\end{table}

\begin{table}[t]
\centering
\caption{Sensitivity of the \(R^2G^3\) algorithm to the upper wake speed ratio threshold.}
\label{tab:sensitivity_upper_ratio}
\begin{tabular}{cccc}
\hline
\textbf{Upper wake speed ratio} & \textbf{Mean} & \textbf{95\% CI (L)} & \textbf{95\% CI (H)} \\
\hline
1.05 & 1.142 & 1.102 & 1.182 \\
1.10 & 0.990 & 0.961 & 1.018 \\
1.15 & 0.949 & 0.922 & 0.977 \\
\hline
\end{tabular}
\end{table}

\bibliography{bibliography}

@article{lundquist2019costs,
  title={Costs and consequences of wind turbine wake effects arising from uncoordinated wind energy development},
  author={Lundquist, Julie K and DuVivier, Katharine K and Kaffine, Daniel and Tomaszewski, Jessica M},
  journal={Nature Energy},
  volume={4},
  number={1},
  pages={26--34},
  year={2019},
  publisher={Nature Publishing Group UK London}
}

@incollection{witha2014high,
  title={High-resolution offshore wake simulations with the LES model PALM},
  author={Witha, Bj{\"o}rn and Steinfeld, Gerald and Heinemann, Detlev},
  booktitle={Wind energy-impact of turbulence},
  pages={175--181},
  year={2014},
  publisher={Springer}
}

@techreport{lissaman1983wake,
  title={Wake structure measurements at the Mod-2 cluster test facility at Goodnoe Hills},
  author={Lissaman, P and Zambrano, TG and Gyatt, GW},
  year={1983},
  institution={AeroVironment, Inc., Pasadena, CA (USA)}
}

@article{lee2017observing,
  title={Observing and simulating wind-turbine wakes during the evening transition},
  author={Lee, Joseph CY and Lundquist, Julie K},
  journal={Boundary-Layer Meteorology},
  volume={164},
  number={3},
  pages={449--474},
  year={2017},
  publisher={Springer}
}

@article{li2025flowformer,
  title={FlowFormer: Toward a foundation model for full-flow-field wind farm wake modeling},
  author={Li, Rui and Zhang, Jincheng and Huang, Yubo and Zhao, Xiaowei},
  journal={Renewable Energy},
  pages={124653},
  year={2025},
  publisher={Elsevier}
}

@article{salim2018microscale,
  title={The microscale obstacle-resolving meteorological model MITRAS v2. 0: model theory},
  author={Salim, Mohamed H and Schl{\"u}nzen, K Heinke and Grawe, David and Boettcher, Marita and Gierisch, Andrea MU and Fock, Bj{\"o}rn H},
  journal={Geoscientific Model Development},
  volume={11},
  number={8},
  pages={3427--3445},
  year={2018},
  publisher={Copernicus Publications G{\"o}ttingen, Germany}
}

@article{xie2015self,
  title={Self-similarity and turbulence characteristics of wind turbine wakes via large-eddy simulation},
  author={Xie, Shengbai and Archer, Cristina},
  journal={Wind Energy},
  volume={18},
  number={10},
  pages={1815--1838},
  year={2015},
  publisher={Wiley Online Library}
}

@article{dorenkamper2015impact,
  title={The impact of stable atmospheric boundary layers on wind-turbine wakes within offshore wind farms},
  author={D{\"o}renk{\"a}mper, Martin and Witha, Bj{\"o}rn and Steinfeld, Gerald and Heinemann, Detlev and K{\"u}hn, Martin},
  journal={Journal of Wind Engineering and Industrial Aerodynamics},
  volume={144},
  pages={146--153},
  year={2015},
  publisher={Elsevier}
}

@article{maronga2020overview,
  title={Overview of the PALM model system 6.0},
  author={Maronga, Bj{\"o}rn and Banzhaf, Sabine and Burmeister, Cornelia and Esch, Thomas and Forkel, Renate and Fr{\"o}hlich, Dominik and Fuka, Vladimir and Gehrke, Katrin Frieda and Geleti{\v{c}}, Jan and Giersch, Sebastian and others},
  journal={Geoscientific Model Development},
  volume={13},
  number={3},
  pages={1335--1372},
  year={2020},
  publisher={Copernicus Publications G{\"o}ttingen, Germany}
}

@techreport{schlunzen2018scientific,
  title={Scientific documentation of the multiscale model system M-SYS},
  author={Schl{\"u}nzen, K Heinke and Boettcher, Marita and Fock, Bj{\"o}rn H and Gierisch, Andrea and Grawe, David and Salim, Mohamed},
  year={2018},
  institution={MEMI Tech. Rep. 4, CEN, Univ. Hambg., 1--153}
}

@article{boettcher2015influence,
  title={Influence of large offshore wind farms on North German climate},
  author={Boettcher, Marita and Hoffmann, Peter and Lenhart, Hermann-J and Schl{\"u}nzen, Heinke and Schoetter, Robert},
  journal={Meteorologische Zeitschrift},
  volume={24},
  pages={465--480},
  year={2015},
  publisher={Borntraeger}
}

@inproceedings{vollmer2015first,
  title={First comparison of LES of an offshore wind turbine wake with dual-Doppler lidar measurements in a German offshore wind farm},
  author={Vollmer, L and Van Dooren, M and Trabucchi, D and Schneemann, J and Steinfeld, G and Witha, B and Trujillo, J and K{\"u}hn, M},
  booktitle={Journal of Physics: Conference Series},
  volume={625},
  number={1},
  pages={012001},
  year={2015},
  organization={IOP Publishing}
}

@inproceedings{witha2014large,
  title={Large-eddy simulation of multiple wakes in offshore wind farms},
  author={Witha, Bj{\"o}rn and Steinfeld, G and D{\"o}renk{\"a}mper, M and Heinemann, D},
  booktitle={Journal of Physics: Conference Series},
  volume={555},
  number={1},
  pages={012108},
  year={2014},
  organization={IOP Publishing}
}

@article{landberg2003wind,
  title={Wind resource estimation—an overview},
  author={Landberg, Lars and Myllerup, Lisbeth and Rathmann, Ole and Petersen, Erik Lundtang and J{\o}rgensen, Bo Hoffmann and Badger, Jake and Mortensen, Niels Gylling},
  journal={Wind Energy: An International Journal for Progress and Applications in Wind Power Conversion Technology},
  volume={6},
  number={3},
  pages={261--271},
  year={2003},
  publisher={Wiley Online Library}
}

@article{djath2018impact,
  title={Impact of atmospheric stability on X-band and C-band synthetic aperture radar imagery of offshore windpark wakes},
  author={Djath, Bughsin and Schulz-Stellenfleth, Johannes and Ca{\~n}adillas, Beatrice},
  journal={Journal of Renewable and Sustainable Energy},
  volume={10},
  number={4},
  year={2018},
  publisher={AIP Publishing}
}

@article{djath2019wind,
  title={Wind speed deficits downstream offshore wind parks--A new automised estimation technique based on satellite synthetic aperture radar data},
  author={Djath, Bughsin and Schulz-Stellenfleth, Johannes},
  journal={Meteorologische Zeitschrift},
  volume={28},
  number={6},
  pages={499--515},
  year={2019},
  publisher={Borntraeger}
}

@article{charnock1955wind,
  title={Wind stress on a water surface},
  author={Charnock, H},
  journal={Quarterly Journal of the Royal Meteorological Society},
  volume={81},
  number={350},
  pages={639--640},
  year={1955},
  publisher={Wiley Online Library}
}

@article{manwell2010wind,
  title={Wind energy explained: theory, design and application},
  author={Manwell, James F and McGowan, Jon G and Rogers, Anthony L},
  year={2010},
  publisher={John Wiley \& Sons}
}

@article{platis2018first,
  title={First in situ evidence of wakes in the far field behind offshore wind farms},
  author={Platis, Andreas and Siedersleben, Simon K and Bange, Jens and Lampert, Astrid and B{\"a}rfuss, Konrad and Hankers, Rudolf and Ca{\~n}adillas, Beatriz and Foreman, Richard and Schulz-Stellenfleth, Johannes and Djath, Bughsin and others},
  journal={Scientific reports},
  volume={8},
  number={1},
  pages={2163},
  year={2018},
  publisher={Nature Publishing Group UK London}
}

@article{lampert2020situ,
  title={In situ airborne measurements of atmospheric and sea surface parameters related to offshore wind parks in the German Bight},
  author={Lampert, Astrid and B{\"a}rfuss, Konrad and Platis, Andreas and Siedersleben, Simon and Djath, Bughsin and Ca{\~n}adillas, Beatriz and Hunger, Robert and Hankers, Rudolf and Bitter, Mark and Feuerle, Thomas and others},
  journal={Earth System Science Data},
  volume={12},
  number={2},
  pages={935--946},
  year={2020},
  publisher={Copernicus GmbH}
}

@article{barfuss2020airborne,
  title={Airborne LiDAR Measurements of Sea Surface Properties in the German Bight},
  author={B{\"a}rfuss, Konrad and Djath, Bughsin and Lampert, Astrid and Schulz-Stellenfleth, Johannes},
  journal={IEEE Transactions on Geoscience and Remote Sensing},
  volume={59},
  number={6},
  pages={4608--4617},
  year={2020},
  publisher={IEEE}
}

@article{christiansen2005wake,
  title={Wake effects of large offshore wind farms identified from satellite SAR},
  author={Christiansen, Merete Bruun and Hasager, Charlotte B},
  journal={Remote Sensing of Environment},
  volume={98},
  number={2-3},
  pages={251--268},
  year={2005},
  publisher={Elsevier}
}

@article{nezhad2021new,
  title={A new methodology for offshore wind speed assessment integrating Sentinel-1, ERA-Interim and in-situ measurement},
  author={Nezhad, M Majidi and Neshat, M and Heydari, A and Razmjoo, A and Piras, G and Garcia, D Astiaso},
  journal={Renewable Energy},
  volume={172},
  pages={1301--1313},
  year={2021},
  publisher={Elsevier}
}

@article{schneemann2020cluster,
  title={Cluster wakes impact on a far-distant offshore wind farm's power},
  author={Schneemann, J{\"o}rge and Rott, Andreas and D{\"o}renk{\"a}mper, Martin and Steinfeld, Gerald and K{\"u}hn, Martin},
  journal={Wind Energy Science},
  volume={5},
  number={1},
  pages={29--49},
  year={2020},
  publisher={Copernicus GmbH}
}

@article{hersbach2010comparison,
  title={Comparison of C-band scatterometer CMOD5. N equivalent neutral winds with ECMWF},
  author={Hersbach, Hans},
  journal={Journal of Atmospheric and Oceanic Technology},
  volume={27},
  number={4},
  pages={721--736},
  year={2010},
  publisher={American Meteorological Society}
}

@article{pettas2021effects,
  title={On the effects of inter-farm interactions at the offshore wind farm Alpha Ventus},
  author={Pettas, Vasilis and Kretschmer, Matthias and Clifton, Andrew and Cheng, Po Wen},
  journal={Wind Energy Science Discussions},
  volume={2021},
  pages={1--25},
  year={2021},
  publisher={G{\"o}ttingen, Germany}
}

@article{yang2014large,
  title={Large-eddy simulation of offshore wind farm},
  author={Yang, Di and Meneveau, Charles and Shen, Lian},
  journal={Physics of Fluids},
  volume={26},
  number={2},
  year={2014},
  publisher={AIP Publishing}
}

@article{hasager2023wind,
  title={Wind Speed-Up in Wind Farm Wakes Quantified From Satellite SAR and Mesoscale Modeling},
  author={Hasager, Charlotte Bay and Imber, James and Fischereit, Jana and Fujita, Aito and Dimitriadou, Krystallia and Badger, Merete},
  journal={Wind Energy},
  volume={27},
  number={11},
  pages={1369--1387},
  year={2024},
  publisher={Wiley Online Library}
}

@article{vanderwende2016simulating,
  title={Simulating effects of a wind-turbine array using LES and RANS},
  author={Vanderwende, Brian J and Kosovi{\'c}, Branko and Lundquist, Julie K and Mirocha, Jeffrey D},
  journal={Journal of Advances in Modeling Earth Systems},
  volume={8},
  number={3},
  pages={1376--1390},
  year={2016},
  publisher={Wiley Online Library}
}

@article{larsen2021case,
  title={A case study of wind farm effects using two wake parameterizations in the Weather Research and Forecasting (WRF) model (V3. 7.1) in the presence of low-level jets},
  author={Lars{\'e}n, Xiaoli G and Fischereit, Jana},
  journal={Geoscientific Model Development},
  volume={14},
  number={6},
  pages={3141--3158},
  year={2021},
  publisher={Copernicus GmbH}
}

@article{pryor2024wind,
  title={Wind shadows impact planning of large offshore wind farms},
  author={Pryor, Sara C and Barthelmie, Rebecca J},
  journal={Applied Energy},
  volume={359},
  pages={122755},
  year={2024},
  publisher={Elsevier}
}

@article{costoya2021climate,
  title={Climate change impacts on the future offshore wind energy resource in China},
  author={Costoya, X and DeCastro, M and Carvalho, D and Feng, Z and G{\'o}mez-Gesteira, M},
  journal={Renewable Energy},
  volume={175},
  pages={731--747},
  year={2021},
  publisher={Elsevier}
}

@article{xu2020proliferation,
  title={Proliferation of offshore wind farms in the North Sea and surrounding waters revealed by satellite image time series},
  author={Xu, Wenxuan and Liu, Yongxue and Wu, Wei and Dong, Yanzhu and Lu, Wanyun and Liu, Yongchao and Zhao, Bingxue and Li, Huiting and Yang, Renfei},
  journal={Renewable and Sustainable Energy Reviews},
  volume={133},
  pages={110167},
  year={2020},
  publisher={Elsevier}
}

@article{hersbach2018era5,
  title={ERA5 hourly data on single levels from 1979 to present},
  author={Hersbach, Hans and Bell, Bill and Berrisford, Paul and Biavati, Gionata and Hor{\'a}nyi, Andr{\'a}s and Mu{\~n}oz Sabater, Joaqu{\'\i}n and Nicolas, Julien and Peubey, Carole and Radu, Raluca and Rozum, Iryna and others},
  journal={Copernicus climate change service (c3s) climate data store (cds)},
  volume={10},
  number={10.24381},
  year={2018},
  publisher={ECMWF Reading, UK}
}

@article{skamarock2019description,
  title={A description of the advanced research WRF version 4},
  author={Skamarock, William C and Klemp, Joseph B and Dudhia, Jimy and Gill, David O and Liu, Zhiquan and Berner, Judith and Wang, Wei and Powers, Jordan G and Duda, Michael G and Barker, Dale M and others},
  journal={NCAR tech. note ncar/tn-556+ str},
  volume={145},
  year={2019},
  publisher={National Center for Atmospheric Research Boulder}
}

@article{howland2022collective,
  title={Collective wind farm operation based on a predictive model increases utility-scale energy production},
  author={Howland, Michael F and Quesada, Jes{\'u}s Bas and Mart{\'\i}nez, Juan Jos{\'e} Pena and Larra{\~n}aga, Felipe Palou and Yadav, Neeraj and Chawla, Jasvipul S and Sivaram, Varun and Dabiri, John O},
  journal={Nature Energy},
  volume={7},
  number={9},
  pages={818--827},
  year={2022},
  publisher={Nature Publishing Group UK London}
}

@article{perera2023challenges,
  title={Challenges resulting from urban density and climate change for the EU energy transition},
  author={Perera, ATD and Javanroodi, Kavan and Mauree, Dasaraden and Nik, Vahid M and Florio, Pietro and Hong, Tianzhen and Chen, Deliang},
  journal={Nature Energy},
  pages={1--16},
  year={2023},
  publisher={Nature Publishing Group UK London}
}

@article{hasager2015using,
  title={Using satellite SAR to characterize the wind flow around offshore wind farms},
  author={Hasager, Charlotte Bay and Vincent, Pauline and Badger, Jake and Badger, Merete and Di Bella, Alessandro and Pe{\~n}a, Alfredo and Husson, Romain and Volker, Patrick JH},
  journal={Energies},
  volume={8},
  number={6},
  pages={5413--5439},
  year={2015},
  publisher={MDPI}
}

@article{jimenez2015mesoscale,
  title={Mesoscale modeling of offshore wind turbine wakes at the wind farm resolving scale: A composite-based analysis with the Weather Research and Forecasting model over Horns Rev},
  author={Jim{\'e}nez, Pedro A and Navarro, Jorge and Palomares, Ana M and Dudhia, Jimy},
  journal={Wind Energy},
  volume={18},
  number={3},
  pages={559--566},
  year={2015},
  publisher={Wiley Online Library}
}

@article{stoelinga2022estimating,
  title={Estimating Long-Range External Wake Losses in Energy Yield and Operational Performance Assessments Using the WRF Wind Farm Parameterization},
  author={Stoelinga, Mark and Sanchez-Gomez, Miguel and Poulos, MS1 Gregory S and Crescenti, Jerry and Renewables, ArcVera},
  journal={Available at: https://arcvera.com/wp-content/uploads/2022/08/ArcVera-White-Paper-Estimating-Long-Range-External-Wake-Losses-WRF-WFP-1.0.pdf},
  year={2022}
}

@article{ma2022jensen,
  title={The Jensen wind farm parameterization},
  author={Ma, Yulong and Archer, Cristina L and Vasel-Be-Hagh, Ahmadreza},
  journal={Wind Energy Science},
  volume={7},
  number={6},
  pages={2407--2431},
  year={2022},
  publisher={Copernicus GmbH}
}

@article{borgers2023mesoscale,
  title={Mesoscale modelling of North Sea wind resources with COSMO-CLM: model evaluation and impact assessment of future wind farm characteristics on cluster-scale wake losses},
  author={Borgers, Ruben and Dirksen, Marieke and Wijnant, Ine L and Stepek, Andrew and Stoffelen, Ad and Akhtar, Naveed and Neirynck, J{\'e}r{\^o}me and Van de Walle, Jonas and Meyers, Johan and van Lipzig, Nicole PM},
  journal={Wind Energy Science Discussions},
  volume={2023},
  pages={1--32},
  year={2023},
  publisher={G{\"o}ttingen, Germany}
}

@article{pryor2021wind,
  title={Wind power production from very large offshore wind farms},
  author={Pryor, Sara C and Barthelmie, Rebecca J and Shepherd, Tristan J},
  journal={Joule},
  volume={5},
  number={10},
  pages={2663--2686},
  year={2021},
  publisher={Elsevier}
}

@article{ahsbahs2018applications,
  title={Applications of satellite winds for the offshore wind farm site Anholt},
  author={Ahsbahs, Tobias and Badger, Merete and Volker, Patrick and Hansen, Kurt S and Hasager, Charlotte B},
  journal={Wind Energy Science},
  volume={3},
  number={2},
  pages={573--588},
  year={2018},
  publisher={Copernicus GmbH}
}

@article{canadillas2020offshore,
  title={Offshore wind farm wake recovery: Airborne measurements and its representation in engineering models},
  author={Ca{\~n}adillas, Beatriz and Foreman, Richard and Barth, Volker and Siedersleben, Simon and Lampert, Astrid and Platis, Andreas and Djath, Bughsin and Schulz-Stellenfleth, Johannes and Bange, Jens and Emeis, Stefan and others},
  journal={Wind Energy},
  volume={23},
  number={5},
  pages={1249--1265},
  year={2020},
  publisher={Wiley Online Library}
}

@article{canadillas2022offshore,
  title={Offshore wind farm cluster wakes as observed by long-range-scanning wind lidar measurements and mesoscale modeling},
  author={Ca{\~n}adillas, Beatriz and Beckenbauer, Maximilian and Trujillo, Juan J and D{\"o}renk{\"a}mper, Martin and Foreman, Richard and Neumann, Thomas and Lampert, Astrid},
  journal={Wind Energy Science},
  volume={7},
  number={3},
  pages={1241--1262},
  year={2022},
  publisher={Copernicus GmbH}
}

@article{finseraas2022gone,
  title={Gone with the Wind? Wind Farm-Induced Wakes and Regulatory Gaps},
  author={Finseraas, Eirik and Herrera Anchustegui, Ignacio and Cheynet, Etienne and Guillermo Gebhardt, Cristian and Reuder, Joachim},
  journal={Available at SSRN: https://ssrn.com/abstract=4294614 or http://dx.doi.org/10.2139/ssrn.4294614},
  year={2022}
}

@article{sebastiani2022evaluation,
  title={Evaluation of the global-blockage effect on power performance through simulations and measurements},
  author={Sebastiani, Alessandro and Pe{\~n}a, Alfredo and Troldborg, Niels and Meyer Forsting, Alexander},
  journal={Wind Energy Science},
  volume={7},
  number={2},
  pages={875--886},
  year={2022},
  publisher={Copernicus GmbH}
}

@article{schneemann2021offshore,
  title={Offshore wind farm global blockage measured with scanning lidar},
  author={Schneemann, J{\"o}rge and Theuer, Frauke and Rott, Andreas and D{\"o}renk{\"a}mper, Martin and K{\"u}hn, Martin},
  journal={Wind Energy Science},
  volume={6},
  number={2},
  pages={521--538},
  year={2021},
  publisher={Copernicus GmbH}
}

@article{lee2008improved,
  title={Improved sigma filter for speckle filtering of SAR imagery},
  author={Lee, Jong-Sen and Wen, Jen-Hung and Ainsworth, Thomas L and Chen, Kun-Shan and Chen, Abel J},
  journal={IEEE Transactions on Geoscience and Remote Sensing},
  volume={47},
  number={1},
  pages={202--213},
  year={2008},
  publisher={IEEE}
}

@article{barthelmie2007modelling,
  title={Modelling and measurements of power losses and turbulence intensity in wind turbine wakes at Middelgrunden offshore wind farm},
  author={Barthelmie, Rebecca Jane and Frandsen, Sten Tron{\ae}s and Nielsen, MN and Pryor, SC and Rethore, P-E and J{\o}rgensen, Hans Ejsing},
  journal={Wind Energy: An International Journal for Progress and Applications in Wind Power Conversion Technology},
  volume={10},
  number={6},
  pages={517--528},
  year={2007},
  publisher={Wiley Online Library}
}

@article{cherp2021national,
  author = {Lei, Yadong and Wang, Zhili and Wang, Deying and Zhang, Xiaoye and Che, Huizheng and Yue, Xu and Tian, Chenguang and Zhong, Junting and Li, Lei and Zhou, Hao and Xu, Yangyang},
  year = {2023},
  month = {06},
  pages = {},
  title = {Co-benefits of carbon neutrality in enhancing and stabilizing solar and wind energy},
  journal = {Nature Climate Change}
}

@article{owda2022wind,
  title={Wind speed variation mapped using SAR before and after commissioning of offshore wind farms},
  author={Owda, Abdalmenem and Badger, Merete},
  journal={Remote Sensing},
  volume={14},
  number={6},
  pages={1464},
  year={2022},
  publisher={MDPI}
}

@article{ali2023assessment,
  title={Assessment of five wind-farm parameterizations in the Weather Research and Forecasting model: A case study of wind farms in the North Sea},
  author={Ali, Karim and Schultz, David M and Revell, Alistair and Stallard, Timothy and Ouro, Pablo},
  journal={Monthly Weather Review},
  year={2023}
}

@article{fitch2012local,
  title={Local and mesoscale impacts of wind farms as parameterized in a mesoscale NWP model},
  author={Fitch, Anna C and Olson, Joseph B and Lundquist, Julie K and Dudhia, Jimy and Gupta, Alok K and Michalakes, John and Barstad, Idar},
  journal={Monthly Weather Review},
  volume={140},
  number={9},
  pages={3017--3038},
  year={2012},
  publisher={American Meteorological Society}
}

@article{siedersleben2018micrometeorological,
  title={Micrometeorological impacts of offshore wind farms as seen in observations and simulations},
  author={Siedersleben, Simon K and Lundquist, Julie K and Platis, Andreas and Bange, Jens and B{\"a}rfuss, Konrad and Lampert, Astrid and Ca{\~n}adillas, Beatriz and Neumann, Tom and Emeis, Stefan},
  journal={Environmental Research Letters},
  volume={13},
  number={12},
  pages={124012},
  year={2018},
  publisher={IOP Publishing}
}

@article{fischereit2022review,
  title={Review of mesoscale wind-farm parametrizations and their applications},
  author={Fischereit, Jana and Brown, Roy and Lars{\'e}n, Xiaoli Guo and Badger, Jake and Hawkes, Graham},
  journal={Boundary-Layer Meteorology},
  volume={182},
  number={2},
  pages={175--224},
  year={2022},
  publisher={Springer}
}

@article{duvivier2020moat,
  title={Moat Mentality: Onshore and Offshore Approaches to Wind Waking},
  author={DuVivier, KK and Mooney, Brendan T},
  journal={Available at SSRN: https://ssrn.com/abstract=2718375 or http://dx.doi.org/10.2139/ssrn.2718375},
  volume={1},
  pages={1},
  year={2020},
  publisher={HeinOnline}
}

@article{ahsbahs2020us,
  title={US East Coast synthetic aperture radar wind atlas for offshore wind energy},
  author={Ahsbahs, Tobias and Maclaurin, Galen and Draxl, Caroline and Jackson, Christopher R and Monaldo, Frank and Badger, Merete},
  journal={Wind Energy Science},
  volume={5},
  number={3},
  pages={1191--1210},
  year={2020},
  publisher={Copernicus GmbH}
}

@article{platis2021evaluation,
  title={Evaluation of a simple analytical model for offshore wind farm wake recovery by in situ data and Weather Research and Forecasting simulations},
  author={Platis, Andreas and Hundhausen, Marie and Mauz, Moritz and Siedersleben, Simon and Lampert, Astrid and B{\"a}rfuss, Konrad and Djath, Bughsin and Schulz-Stellenfleth, Johannes and Canadillas, Beatriz and Neumann, Thomas and others},
  journal={Wind Energy},
  volume={24},
  number={3},
  pages={212--228},
  year={2021},
  publisher={Wiley Online Library}
}

\end{document}